\documentclass[12pt]{iopart}

\usepackage{amsfonts}
\usepackage{graphicx}
\usepackage{subfigure}
\usepackage{dcolumn}
\usepackage{bm}
\usepackage{booktabs}\usepackage[utf8]{inputenc}
\usepackage{multirow}
\usepackage{graphicx,graphics,dcolumn,booktabs,bm}
\usepackage{longtable,lscape}
\usepackage{txfonts}
\usepackage{overpic}
\usepackage{amssymb}
\usepackage{indentfirst}
\usepackage{epsfig}
\usepackage{feynmf}   
\usepackage{epstopdf}   
\usepackage{slashed}  
\usepackage{color}
\usepackage[section]{placeins}
\usepackage{iopams}
\def\cbar{\overline{c}}

\def\taubar{\overline{\tau}}

\def\Leff{\mathcal{L}_{\rm eff}}

\def\A{\mathcal{A}}
\def\O{\mathcal{O}}

\def\th{{\theta}}

\begin{document}
\title{Scrutinizing New Physics in Semi-leptonic $B_{c}\rightarrow J/\psi \tau\nu$ Decay}

\author{Ru-Ying Tang$^{1,2}$, Zhuo-Ran Huang$^3$, Cai-Dian L\"u$^{1,2}$ and Ruilin Zhu$^4$}

\address{$^1$Institute of High Energy Physics, Chinese Academy
of Sciences, Beijing 100049, China}
\address{$^2$School of Physics, University of Chinese Academy of Sciences, Beijing 100049, China}
\address{$^3$Asia Pacific Center for Theoretical Physics, Pohang, 37673, Korea}
\address{$^4$Department of Physics and Institute of Theoretical Physics, Nanjing Normal University, Nanjing 210023, China}
\ead{tangruying@ihep.ac.cn, zhuoran.huang@apctp.org, lucd@ihep.ac.cn and rlzhu@njnu.edu.cn}
\vspace{10pt}

\begin{abstract}
We perform a global analysis of the $b\to c\tau\nu$ data using the recent lattice results on the $B_c\to J/\psi$ vector and axial-vector form factors. To explore the effects from the tensor operator of new physics beyond the Standard Model, we determine the tensor form factors by using the non-relativistic QCD (NRQCD) relations between tensor and (axial-)vector form factors. Based on the lattice+NRQCD form factors, we fit the Wilson coefficients and the new physics couplings in $R_2$, $S_1$ and $U_1$ leptoquark models by including the recently measured $R(\Lambda_c)$ and imposing the relaxed constraint $\mathcal B(B_c\to J/\psi)<30\%$ in light of the recent studies on LEP1 data and $B_c$ lifetime. We give predictions for the experimental observables including $R(J/\psi)$, $P_\tau(J/\psi)$, $F_L(J/\psi)$ and $\mathcal A_{FB}(J/\psi)$ as well as their $q^2$ distribution in new physics scenarios/models. Our results suggest that the longitudinal $\tau$ polarization fraction $P_\tau(J/\psi)$ and the forward-backward asymmetry $\mathcal A_{FB}(J/\psi)$ are useful for testing the $R_2$ leptoquark model.
\end{abstract}

%
%
%
%
%
\section{Introduction}\label{sec:Intro}
Since the BaBar Collaboration reported for the first time the discrepancy between the measurement and the Standard Model (SM) prediction of the lepton flavour universality (LFU) ratio $R(D^{(*)})$~\cite{Lees:2012xj,Lees:2013uzd}, Belle and LHCb have also carried out several measurements   on $R(D)$ and/or $R(D^*)$ based on different datasets and tagging methods~\cite{Huschle:2015rga,Sato:2016svk,Hirose:2016wfn,Abdesselam:2019dgh,Aaij:2015yra,Aaij:2017uff,Aaij:2017deq,Aaij:2017tyk,Adamczyk:2019wyt,Abdesselam:2019wbt}. At present, the SM  predictions for $R(D^{(*)})$ \cite{Bigi:2016mdz,Bernlochner:2017jka,Bigi:2017jbd,Jaiswal:2017rve} still have a $3$-$4\sigma$ deviation from the experimental average~\cite{HFLAV:2019otj}. Therefore, to further test the lepton flavour universality, experimental and theoretical study of $b\to c\ell\nu$ channels complementary to $B\to D^{(*)}\ell\nu$ will play a pivotal role. One of such channels, $B_c \to J/\psi\ell\nu$ has been investigated by LHCb~\cite{Aaij:2017tyk}.  The first measurement of the LFU ratio
\begin{eqnarray}
	R(J/\psi)=\frac{\mathcal B(B_{c}\to J/\psi\tau\nu)}{\mathcal B(B_{c}\to J/\psi\mu\nu)},\label{eq:Rjpsi}
\end{eqnarray}
was $R(J/\psi)=0.71\pm 0.17\pm 0.18$ \cite{Aaij:2017tyk} which showed a deviation from the expectation in the SM of $\sim2\sigma$ depending on the determination of $B_c\to J/\psi$ form factors\footnote{See \cite{Huang:2018nnq,Tran:2018kuv} for summaries of the theoretical calculation of $B_c\to J/\psi$ form factors.}. Although the uncertainty of the experimental data is still large, $R(J/\psi)$ can provide extra possibility to investigate new physics  effects in the $b\to c\tau\nu$ transition~\cite{Huang:2018nnq,Watanabe:2017mip,Tran:2018kuv,Azatov:2018knx,Dutta:2017xmj,Hu:2018veh,Bhattacharya:2018kig,Shi:2019gxi,Asadi:2019xrc,Murgui:2019czp,Gomez:2019xfw}.

On the theoretical side, the calculation of the $B_c \to J/\psi$ form factors is crucial for the determination of $R(J/\psi)$, which had been a sticky problem before the recent lattice QCD result was available~\cite{Harrison:2020}. People performed calculation using different methods such as non-relativistic QCD (NRQCD) effective theory~\cite{Zhu:2017lqu,Shen:2014msa,Shen:2021dat,Zhu:2017lwi}, perturbative QCD ~\cite{Wen-Fei:2013uea,Rui:2016opu}, QCD sum rules~\cite{Kiselev:2002vz,Leljak:2019eyw}, QCD light-cone sum rules~\cite{Fu:2018vap,Zhong:2018exo} and different types of quark models~\cite{Wang:2008xt,Hernandez:2006gt,Tran:2018kuv}. Apart from the model calculations, lattice QCD has given an updated result on $R(J/\psi)$~\cite{Harrison:2020nrv} based on the form factor calculation~\cite{Harrison:2020}. With the help of these lattice results, predictions for more observables as well as analysis of new physics for the $B_c \to J/\psi\tau\nu$ decay can also be provided.

However, to perform an analysis of the new physics effects, we need not only the vector and axial-vector form factors which are given by lattice calculation~\cite{Harrison:2020}, but also the tensor form factors which are absent in \cite{Harrison:2020}. As a consequence, it is still necessary to estimate the tensor form factors using theoretical methods other than lattice QCD, and in the meantime, it should also be important to use the information from lattice QCD calculation as much as possible. NRQCD is such a method that fulfill both conditions, because it is an effective theory derived from  QCD valid for doubly heavy quark systems and provide relations between the tensor form factors and the (axial-) vector form factors calculated by lattice QCD, which allows us to obtain predictions for the tensor form factors based on the (axial-) vector form factors in lattice QCD.

In this work, we try to obtain the lattice+NRQCD tensor form factors, and use them to calculate the observables for the $B_c\to J/\psi\tau\nu$ decay in the SM and the new physics scenarios/models, including the lepton flavour universality ratio $R(J/\psi)$, the longitudinal $\tau$ polarization fraction $P_{\tau}(J/\psi)$, the longitudinal $J/\psi$ polarization fraction $F_L(J/\psi)$ and the forward-backward asymmetry of the $\tau$ lepton $\mathcal{A}_{FB}(J/\psi)$. To estimate the size of new physics effects, we perform a global fit of the Wilson coefficients to the existing $b\to c\tau\nu$ data including the LFU ratio $R(\Lambda_c)$ recently measured by LHCb~\cite{LHCb:2022piu} for $\Lambda_b\to\Lambda_c\tau\nu$ decay \cite{Mu:2019bin,Hu:2020axt,Becirevic:2020nmb,Bernlochner:2018bfn,Shivashankara:2015cta,Cabarcas:2022mfa,Ray:2018hrx,Duan:2022uzm}. We study both the model-independent scenarios and the leptoquark models including $R_2$, $S_1$ and $U_1$, which are expected to be able to explain the $b\to c\tau\nu$ anomalies. The fit results update the results in \cite{Cheung:2020sbq} by using the lattice + NRQCD form factors for $B_c\to J/\psi$ and incorporating $R(\Lambda_c)$, and we also use best-fit results to make predictions for various observables in the presence of new physics.

The paper is organized in the following way: in the next section we introduce the basic framework of weak effective theory and operator basis. In Section~\ref{sec:Bcff} and \ref{sec:tensorfit} we present the NRQCD relations of $B_c\to J/\psi$ form factors and the fit of the tensor form factors, respectively. In Section~\ref{sec:CWC} and \ref{sec:PRE} we present the results of the global fit of the Wilson coefficients and the predictions for $B_c\to J/\psi\tau\nu$ observables, respectively. Finally, we summarize the work in Section~\ref{sec:SUM}.
\section{Effective Four-Fermion Interactions and Operator Basis}\label{sec:EFT}
In the SM, the $B_{c}\to J/\psi\tau\nu$ decay via $b\to c\tau\nu$ can be described by the left-handed vector four-fermion (V-A) interaction as an effective theory. If one wants to study new physics effects, the effective Lagrangian should be extended to contain the full basis of four-fermion operators. In this work we take the assumption that neutrinos are left-handed and their flavors are not differentiated, therefore the effective Lagrangian can be written as
\begin{equation}
   \Leff =- \frac{4G_F}{ \sqrt{2}} V_{cb}\left[ (1 + C_{V_1})\O_{V_1} + C_{V_2}\O_{V_2} + C_{S_1}\O_{S_1} + C_{S_2}\O_{S_2} + C_T\O_T \right] + \rm{h.c.} \,,
      \label{eq:lag}
\end{equation}
where the four-fermion operators are defined as
\begin{eqnarray}
 &\O_{S_1} =  (\cbar_L b_R)(\taubar_R \nu_{L}) \,, \,\,\, \O_{S_2} =  (\cbar_R b_L)(\taubar_R \nu_{L}) \,, \,\,\, \nonumber\\
   & \O_{V_1} = (\cbar_L \gamma^\mu b_L)(\taubar_L \gamma_\mu \nu_{L}) \,,  \,\,\,
     \O_{V_2} = (\cbar_R \gamma^\mu b_R)(\taubar_L \gamma_\mu \nu_{L}) \,,  \,\,\, \nonumber\\
  & \O_T =  (\cbar_R \sigma^{\mu\nu} b_L)(\taubar_R \sigma_{\mu\nu} \nu_{L}) \,,
   \label{eq:operators}
\end{eqnarray}
and $C_X$ ($X=S_1$, $S_2$, $V_1$, $V_2$, and $T$) are the corresponding Wilson coefficients, which are non-vanishing only in the presence of new physics. With the effective Lagrangian given in Equation~(\ref{eq:lag}), the differential decay rates for $B_{c}\to J/\psi\tau\nu$ can be expressed by the following helicity amplitudes in terms of the hadronic matrix elements~\cite{Tanaka:2012nw,Sakaki:2013bfa}:
\begin{eqnarray}
 H^{\lambda_{J/\psi}}_{V_{1,2,\lambda}}(q^{2})&=&\varepsilon^{\ast}_{\mu}(\lambda)\langle J/\psi(\lambda_{J/\psi})|\bar c\gamma^{\mu}(1\mp\gamma_{5})b|B_{c}\rangle \, ,\nonumber\\
  H^{\lambda_{J/\psi}}_{S_{1,2,\lambda}}(q^{2})&=&\langle J/\psi(\lambda_{J/\psi})|\bar c(1\pm\gamma_{5})b|B_{c}\rangle \, ,\nonumber\\
  H^{\lambda_{J/\psi}}_{T,\lambda\lambda^{\prime}}(q^{2})&=&- H^{\lambda_{J/\psi}}_{T,\lambda^{\prime}\lambda}(q^{2})= \varepsilon^{\ast}_{\mu}(\lambda)\varepsilon^{\ast}_{\nu}(\lambda^{\prime})\langle J/\psi(\lambda_{J/\psi})|\bar c\sigma^{\mu\nu}(1-\gamma_{5})b|B_{c}\rangle \, ,
 \label{eq:Helicity}
\end{eqnarray}
where $\lambda_{J/\psi}=0,\pm 1$ and $\lambda,\lambda'=0,\pm 1,t$ represent the helicity of $J/\psi$ and the virtual $W$ boson, respectively. The hadronic matrix elements in Equation~(\ref{eq:Helicity}) encoding the long-distance effects can be expressed by the hadronic form factors, which then allow one to express the observables of interest in terms of the invariant form factors. The detailed analytic expressions are presented in ~\cite{Watanabe:2017mip}, which we use for the phenomenological study.
\section{$B_c \to J/\psi$ form factors and NRQCD relations}\label{sec:Bcff}
The $B_c\to J/\psi$ hadronic matrix elements of $\bar c \Gamma b$ local currents can be expressed by the $B_c\to J/\psi$ invariant form factors as follows~\cite{Wirbel:1985}:
\begin{eqnarray}
	\langle J/\psi(p,\varepsilon^{*})\vert \bar c \gamma^\mu b\vert
	B_{c}(P)\rangle&=&\frac{2V^{ J/\psi}(q^{2})}{m_{B_{c}}+m_{J/\psi}}\epsilon^{\mu\nu\rho\sigma}
	\varepsilon_{\nu}^{*}p_{\rho}P_{\sigma}\,,
	\\
	\langle J/\psi(p,\varepsilon^{*})\vert \bar c \gamma^\mu \gamma^5 b\vert B_{c}(P)\rangle&=&-i\left[2
	m_{J/\psi}A^{ J/\psi}_{0}(q^{2})\frac{\varepsilon^{*}\cdot q}{q^{2}}q^{\mu}\right. \nonumber\\
    &&\left.+(m_{B_{c}}+m_{J/\psi})A^{ J/\psi}_{1}(q^{2})
	(\varepsilon^{*\mu}-\frac{\varepsilon^{*}\cdot q}{q^{2}} q^{\mu})\right.\nonumber\\
	&&\left.-A^{ J/\psi}_{2}(q^{2})\frac{\varepsilon^{*}\cdot
		q}{m_{B_{c}}+m_{J/\psi}}(
	P^{\mu}+p^{\mu}-\frac{m_{B_{c}}^{2}-m_{J/\psi}^{2}}{q^{2}}q^{\mu})\right]\,,\nonumber\\
    \\
	\langle J/\psi(p,\varepsilon^{*})\vert \bar c \sigma^{\mu\nu}q_\nu b\vert
	B_{c}(P)\rangle&=&2i
	T_1^{ J/\psi}(q^{2})\epsilon^{\mu\nu\rho\sigma}
	\varepsilon_{\nu}^{*}p_{\rho}P_{\sigma}\,,
	\\
	\langle J/\psi(p,\varepsilon^{*})\vert \bar c \sigma^{\mu\nu}\gamma^5 q_\nu   b\vert B_{c}(P)\rangle&=&T_2^{ J/\psi}(q^{2})\left((m_{B_{c}}^{2}-m_{J/\psi}^{2})\varepsilon^{*\mu}-\varepsilon^{*}\cdot q(P^{\mu}+p^{\mu})\right)
	\nonumber\\&&+T^{ J/\psi}_{3}(q^{2})\varepsilon^{*}\cdot
	q\left[ q^{\mu}-\frac{q^{2}}{m_{B_{c}}^{2}-m_{J/\psi}^{2}}(P^{\mu}+p^{\mu})\right]
	\,,
\end{eqnarray}
where $m_{B_c}$ and $m_{J/\psi}$ are the masses of $B_c$ and $J/\psi$ meson, respectively. $\epsilon^*$ is the polarization vector of the outgoing $J/\psi$ meson. $P^\mu$ and $p^\mu$ are the four momenta of $B_c$ and $J/\psi$, respectively. The momentum transfer $q^\mu$ is defined as $q^\mu=P^\mu-p^\mu$. Besides, the conventions $\sigma_{\mu\nu}={i\over 2} [\gamma_\mu,\gamma_\nu]$ and $\epsilon^{0123}=1$ are used.

Since $B_c$ and $J/\psi$ are both composed of two heavy quarks, we can use NRQCD to relate different form factors. In NRQCD, both the $B_c$ meson and $J/\psi$ can be treated as non-relativistic bound states and the theory  is expanded  by the power of relative velocity of the quark in the heavy meson. The decay amplitudes for $B_c\to J/\psi$ can be factorized as the short-distance Wilson coefficients and the long-distance matrix elements. The leading order results for the form factors in NRQCD are~\cite{Qiao:2012vt,Zhu:2017lqu,Shen:2021dat}
\begin{eqnarray}\label{NRQCD}
	V^{J/\psi,\,LO}(r,y)
	&=&\frac{8 \sqrt{2} \pi \left(r+1\right)^{3/2} (3 r+1) C_F \alpha _s}{r^{3/2} m_b^3 N_c \left((r-1)^2-y^2\right)^2}\langle J/\psi|\psi_{c}^{\dagger} \bm{\sigma}\chi_{c}| 0\rangle \langle 0|\psi_{b}^{\dagger} \chi_{c}| B_c\rangle\,,\\
	A^{J/\psi,\,LO}_{0}(r,y)
	&=&\frac{8 \sqrt{2} \pi
		(r+1)^{5/2} C_F \alpha _s}{r^{3/2} m_b^3 N_c \left((r-1)^2-y^2\right)^2}\langle J/\psi|\psi_{c}^{\dagger}\bm{\sigma} \chi_{c}| 0\rangle \langle 0|\psi_{b}^{\dagger} \chi_{c}| B_c\rangle\,,\\
	A^{J/\psi,\,LO}_{1}(r,y)
	&=&\frac{8 \sqrt{2} \pi   \sqrt{r+1}
		\left(4 r^3+5 r^2-(2 r+1) y^2+6 r+1\right)C_F  \alpha _s}{r^{3/2} (3 r+1) m_b^3 N_c \left((r-1)^2-y^2\right)^2}\nonumber\\&&\times\langle J/\psi|\psi_{c}^{\dagger}\bm{\sigma} \chi_{c}| 0\rangle \langle 0|\psi_{b}^{\dagger} \chi_{c}| B_c\rangle\,,~~~~\\
	A^{J/\psi,\,LO}_{2}(r,y)
	&=&\frac{8 \sqrt{2} \pi
		\sqrt{r+1} (3 r+1) C_F \alpha _s}{r^{3/2} m_b^3 N_c \left((r-1)^2-y^2\right)^2}\langle J/\psi|\psi_{c}^{\dagger}\bm{\sigma} \chi_{c}| 0\rangle \langle 0|\psi_{b}^{\dagger} \chi_{c}| B_c\rangle\,,\\
	T^{J/\psi,\,LO}_{1}(r,y)
	&=&\frac{2 \sqrt{2} \pi   \sqrt{r+1} \left(5
		r^2+6 r-y^2+5\right)C_F  \alpha _s}{r^{3/2} m_b^3 N_c \left((r-1)^2-y^2\right)^2}\langle J/\psi|\psi_{c}^{\dagger}\bm{\sigma} \chi_{c}| 0\rangle \langle 0|\psi_{b}^{\dagger} \chi_{c}| B_c\rangle\,,\nonumber\\ \\
	T^{J/\psi,\,LO}_{2}(r,y)
	&=&\frac{2 \sqrt{2} \pi  \sqrt{r+1} \left(15
		r^4+8 r^3-2 r^2 \left(3 y^2+1\right)-16 r-y^4+6 y^2-5\right) C_F \alpha _s}{(r-1) r^{3/2} (3 r+1) m_b^3 N_c \left((r-1)^2-y^2\right)^2}\nonumber\\&&\times\langle J/\psi|\psi_{c}^{\dagger}\bm{\sigma} \chi_{c}| 0\rangle \langle 0|\psi_{b}^{\dagger} \chi_{c}| B_c\rangle\,,\\
	T^{J/\psi,\,LO}_{3}(r,y)
	&=&-\frac{2 \sqrt{2} \pi
		\sqrt{r+1}\left(3 r^2+2 r+y^2-5\right) C_F \alpha _s}{r^{3/2} m_b^3 N_c \left((r-1)^2-y^2\right)^2}\langle J/\psi|\psi_{c}^{\dagger}\bm{\sigma} \chi_{c}| 0\rangle \langle 0|\psi_{b}^{\dagger} \chi_{c}| B_c\rangle\,,\nonumber\\
\end{eqnarray}
where $r=m_c/m_b$ and $y=\sqrt{q^2/m_b^2}$.   $\langle J/\psi|\psi_{c}^{\dagger}\bm{\sigma} \chi_{c}| 0\rangle$ and  $\langle 0|\psi_{b}^{\dagger} \chi_{c}| B_c\rangle$ are the NRQCD long-distance matrix elements. Because different form factors have the same combination of long-distance matrix elements, their ratios are independent of the long-distance effects at leading power.

We also consider relativistic corrections from the relative velocity of the quark in the meson. We use the formulae in \cite{Shen:2021dat} and keep $\mathcal{O}(1/r)$ corrections. The next-to-leading order of relativistic corrections from the relative velocity of the quark in $J/\psi$ contribute to various form factors as
\begin{eqnarray}\label{NRQCDRC1}
	K^{RC1}(V^{J/\psi}(r,y))
	&=&\frac{|\bm{k}_{c}|^{2}}{3m_{b}^{2}r}\left(\frac{1}{r}+\frac{y^{2}-27}{y^{2}-1}\right)\,,\\
	K^{RC1}({A_0}^{J/\psi}(r,y))
	&=&\frac{|\bm{k}_{c}|^{2}}{3m_{b}^{2}r}\left(\frac{5}{2r}-\frac{2y^{2}+24}{y^{2}-1}\right)\,,\\
	K^{RC1}(A^{J/\psi}_{1}(r,y))
	&=&\frac{|\bm{k}_{c}|^{2}}{3m_{b}^{2}r}\left(\frac{1}{r}-\frac{y^{2}+53}{2y^{2}-2}\right)\,,\\
	K^{RC1}(A^{J/\psi}_{2}(r,y))
	&=&\frac{|\bm{k}_{c}|^{2}}{3m_{b}^{2}r}\left(\frac{1}{r}+\frac{y^{2}-41}{2y^{2}-2}\right)\,,\\
	K^{RC1}(T^{J/\psi}_{1}(r,y))
	&=&-\frac{|\bm{k}_{c}|^{2}}{3m_{b}^{2}r(y^{2}-5)}\left(\frac{3y^{2}+5}{2r}+\frac{32y^{4}-236y^{2}+620}{(y^{2}-1)(y^{2}-5)}\right)\,,\\
	K^{RC1}(T^{J/\psi}_{2}(r,y))
	&=&-\frac{|\bm{k}_{c}|^{2}}{3m_{b}^{2}r(y^{2}-5)}\left(\frac{3y^{2}+5}{2r}+\frac{8y^{4}-196y^{2}+620}{(y^{2}-1)(y^{2}-5)}\right)\,,\\
	K^{RC1}(T^{J/\psi}_{3}(r,y))
	&=&-\frac{|\bm{k}_{c}|^{2}}{3m_{b}^{2}r(y^{2}-5)}\left(\frac{3y^{2}+5}{2r}+\frac{8y^{4}-148y^{2}+460}{(y^{2}-1)(y^{2}-5)}\right)\,,
\end{eqnarray}
where $\bm{k}_{c}=m_{c}\bm{v}_{c}/2$ is half of the relative momentum of quarks in $J/\psi$, and we use the value of $\bm{v}_{c}$ in \cite{WANG:2015NRQCD}.

Similarly, the next-to-leading order relativistic corrections from $B_{c}$ meson contribute to form factors as
\begin{eqnarray}\label{NRQCDRC2}
	K^{RC2}(V^{J/\psi}(r,y))
	&=&\frac{|\bm{k}_{bc}|^{2}}{12m_{b}^{2}r}\left(-\frac{7}{2r}+13\right)\,,\\
	K^{RC2}(A^{J/\psi}_{0}(r,y))
	&=&\frac{|\bm{k}_{bc}|^{2}}{12m_{b}^{2}r}\left(\frac{5}{2r}+\frac{16y^{2}-32}{y^{2}-1}\right)\,,\\
	K^{RC2}(A^{J/\psi}_{1}(r,y))
	&=&\frac{|\bm{k}_{bc}|^{2}}{12m_{b}^{2}r}\left(-\frac{7}{2r}+\frac{11y^{2}-31}{y^{2}-1}\right)\,,\\
	K^{RC2}(A^{J/\psi}_{2}(r,y))
	&=&\frac{|\bm{k}_{bc}|^{2}}{12m_{b}^{2}r}\left(-\frac{7}{2r}+\frac{15y^{2}-7}{y^{2}-1}\right)\,,\\
	K^{RC2}(T^{J/\psi}_{1}(r,y))
	&=&\frac{|\bm{k}_{bc}|^{2}}{12m_{b}^{2}r(y^{2}-5)}\left(\frac{-11y^{2}+7}{2r}-\frac{y^{6}+105y^{4}-581y^{2}+731}{(y^{2}-1)(y^{2}-5)}\right)\,,\nonumber\\ \\
	K^{RC2}(T^{J/\psi}_{2}(r,y))
	&=&\frac{|\bm{k}_{bc}|^{2}}{12m_{b}^{2}r(y^{2}-5)}\left(\frac{-11y^{2}+7}{2r}-\frac{y^{6}+85y^{4}-733y^{2}+1031}{(y^{2}-1)(y^{2}-5)}\right)\,,\nonumber\\ \\
	K^{RC2}(T^{J/\psi}_{3}(r,y))
	&=&\frac{|\bm{k}_{bc}|^{2}}{12m_{b}^{2}r(y^{2}-5)}\left(\frac{-11y^{2}+7}{2r}-\frac{y^{6}+73^{4}-613^{2}+923}{(y^{2}-1)(y^{2}-5)}\right)\,,
\end{eqnarray}
where $\bm{k}_{bc}=m_{b}m_{c}\bm{v}_{bc}/(m_{b}+m_{c})$ is the reduced relative momentum of quarks in $B_{c}$, and we also use the value of $\bm{v}_{bc}$ in \cite{WANG:2015NRQCD}. With these relativistic corrections, form factors in NRQCD are expressed as
\begin{equation}\label{NRQCDALL}
	F_{NRQCD}=F_{NRQCD}^{LO}(1+K_{NRQCD}^{RC1}+K_{NRQCD}^{RC2})\,.
\end{equation}
\section{Fit of tensor form factors}\label{sec:tensorfit}
As mentioned in the previous section, we utilize the lattice QCD results and NRQCD relations to determine the $B_c\to J/\psi$ tensor form factors. In the lattice calculation~\cite{Harrison:2020}, the $q^2$ dependence of the form factors present in the SM are parameterized as
\begin{equation}\label{latparame}
	F^{J/\psi}(q^{2})
	=\frac{1}{P_{F}(q^{2})}\sum_{n=0}^{3}a_{F,n}z^{n}\,,(F=V,\,A_{0},\,A_{1},\,A_{2})\,,
\end{equation}
where the Blaschke factors corresponding to different $J^P$ quantum numbers ($1^-$ for $V$, $0^-$ for $A_0$ and $1^+$ for $A_1$ and $A_2$) of the subthreshold $B_c$ poles are defined as~\cite{Harrison:2020}
\begin{equation}
P_{F}(q^{2})=\prod_{B_c\,\textrm{poles}}\frac{\sqrt{t_+-q^2}-\sqrt{t_+-m_{B_c}^2}}{\sqrt{t_+-q^2}-\sqrt{t_++m_{B_c}^2}}\,.
\end{equation}
Equation~(\ref{latparame}) can be considered as a simplified version of the Boyd, Grinstein and Lebed  parametrization \cite{Boyd:1997kz} of form factors without the outer functions. Central values, uncertainties and correlations of $a_{F,n}$ are given in \cite{Harrison:2020}, which we use   to generate input data for the tensor form factors in the large recoil region, through the following relation:
\begin{equation}
	T^{J/\psi}_{i,\,DATA}(q^{2}_{k})=\frac{T_{i,\,NRQCD}(q^{2}_{k})}{F_{NRQCD}(q^{2}_{k})}F^{J/\psi}(q^{2}_{k})\,,\,(q^{2}_{1}=0\,,k=1,2,...,m)
\end{equation}
where we take $q^2_k$ at equal spacing with $q^2_{max}=3$ ($m=4$) to ensure the validity of the NRQCD relations~\cite{Zhu:2017lqu}. Equation~(\ref{NRQCDALL}) is used for NRQCD form factors in both the numerator and the denominator, thus the long-distance matrix elements get explicitly cancelled.
{\renewcommand{\arraystretch}{1.2} 
\begin{table}
	\centering
	\caption{Fitted values of $B_{c}\to J/\psi$ tensor form factors.}
\begin{small}
	\begin{tabular}{c c c c c}
		\hline
		\hline
		& $a_{0}$  &$a_{1}$ & $a_{2}$&$a_{3}$ \\
		\hline
		$T_{1}$&0.0640(10)&$-0.432(26)$&0.761(195)&$-1.263(438)$\\
		$T_{2}$&0.0528(8)&$-0.152(22)$&$-0.088(129)$&$-2.358(1201)$\\
		$T_{3}$&0.0351(5)&$-0.189(14)$&0.054(103)&0.042(316)\\
		\hline
		\hline
	\end{tabular}
\end{small}
	\label{tab:fitresult}
\end{table}}
We use all the (axial)-vector form factors and consider full correlations among them in generating the input data in the large hadronic recoil region. To extrapolate the form factors to the small recoil region, we use the same parametrization for the tensor form factors as in Equation~(\ref{latparame}) for the (axial-)vector form factors. The tensor form factors read
\begin{equation}\label{parame}
	T^{J/\psi}_{i}(q^{2})
	=\frac{1}{P(q^{2})}\sum_{n=0}^{3}a_{i,n}z^{n}\,,(i=1,\,2,\,3)\,,
\end{equation}
where $a_{i,n}$ are the $z$-expansion parameters to be fitted. We adopt the masses of the tensor $B_{c}$ resonances ($1^+$ for $T_1$, and $1^-$ for $T_2$ and $T_3$) in ref.~\cite{Eichten:1994}. We utilize the least square method for the fit, and the $\chi^{2}$ function for each tensor form factor is defined as
\begin{equation}\label{fit} \chi^{2}=\sum_{l,k}\left(T^{J/\psi}_{i}(q^{2}_{l})-T^{J/\psi}_{i,\,DATA}(q^{2}_{l})\right)\left(V_{DATA}^{-1}\right)_{l,k}\left(T^{J/\psi}_{i}(q^{2}_{k})-T^{J/\psi}_{i,\,DATA}(q^{2}_{k})\right)\,.
\end{equation}
By minimizing the $\chi^{2}$ function, we obtain the fitted results listed in Table~\ref{tab:fitresult} and Table~\ref{tab:fitcorrelations}. With these results on tensor form factors and the lattice results on (axial-)vector form factors, we study constraints on new physics and predictions for observables in the following two sections.
\begin{table*}
\centering
\caption{Fitted correlations of $a_{n}$ in $B_{c}\to J/\psi$ tensor form factors.}
\begin{small}
\begin{tabular}{c c c c c|| c c c c c}
	\hline
	\hline
	$T_{1}$& $a_{0}$  &$a_{1}$ & $a_{2}$&$a_{3}$&$\ T_{2}$& $a_{0}$  &$a_{1}$ & $a_{2}$&$a_{3}$\\
	\hline
	$a_{0}$&1& 0.0333 &$  -0.0292 $ &$  -0.0335 $&
	$\ a_{0}$&1& 0.1804& 0.0621 &$  -0.0340 $\\
	$a_{1}$&0.0333 & 1 & $ -0.3543  $& 0.0606&
	$\ a_{1}$&0.1804 & 1 & 0.2624 & 0.0692\\
	$a_{2}$&$ -0.0292 $&$  -0.3543 $ & 1&$  -0.0465$&
	$\ a_{2}$&0.0621 & 0.2624 & 1 & 0.6382\\
	$a_{3}$& $ -0.0335  $& 0.0606&$  -0.0465 $& 1& 
	$\ a_{3}$&$ -0.0340 $& 0.0692 & 0.6382 & 1\\
	\hline
	\hline
\end{tabular}
\begin{tabular}{c c c c c}
	\hline
	\hline
	$\ T_{3}$& $a_{0}$  &$a_{1}$ & $a_{2}$&$a_{3}$ \\
	\hline
	$\ a_{0}$&1&0.0859& $ -0.0326 $ &$  -0.0337 $\\
	$\ a_{1}$&0.0859 & 1 &$  -0.1956 $& $ -0.1103 $\\
	$\ a_{2}$& $ -0.0326 $ &$  -0.1956 $& 1 & 0.3369\\
	$\ a_{3}$&$ -0.0337 $&$  -0.1103  $& 0.3369 & 1\\
	\hline
	\hline
\end{tabular}
\end{small}
\label{tab:fitcorrelations}
\end{table*}
\section{experimental constraints on the Wilson Coefficients}\label{sec:CWC}
In this section, we use the experimental data to analyze the allowed parameter space of new physics scenarios/models. Our analyses include two parts: the model-independent analysis and the analysis of the leptoquark models $R_2$, $S_1$ and $U_1$. First we perform the fit of the Wilson coefficients or the leptoquark couplings to the experimental measurements considering the constraint from $B_c\to\tau\nu$ decay. The stringent constraint $\mathcal B(B_c\to\tau\nu)<10\%$ was obtained from the LEP1 data taken at Z peak~\cite{Akeroyd:2017mhr}. It was found in \cite{Cheung:2020sbq} that the $R_2$ leptoquark model is in tension with such a stringent constraint because the dominant contribution from $R_2$ is the scalar contribution $\O_{S_2}$. However, it was found recently from the reconsideration of both the LEP1 data~\cite{Blanke:2019qrx} and $B_c$ lifetime calculation~\cite{Aebischer:2021ilm} that such a constraint on $\mathcal B(B_c\to\tau\nu)$ can be significantly relaxed, therefore we use the less restrictive $\mathcal B(B_c\to\tau\nu)<30\%$~\cite{Alonso:2016oyd} in this work.
\begin{table*}
\centering
\caption{Experimental data used in the fit.}
\begin{scriptsize}
\begin{tabular}{c c c c c c c c}
\hline
\hline
    & $R(D)$ & $R(D^*)$  &$P_\tau(D^*)$ & $F_L^{D^{*}}$&$R(J/\psi)$& $R(\Lambda_c)$ &Correlation   \\
\hline
BaBar~\cite{Lees:2012xj,Lees:2013uzd}       &$0.440(58)(42)$    &$0.332(24)(18)$     &$-$&$-$&$-$&-&$-0.27$ \\
Belle~\cite{Huschle:2015rga}       &$0.375(64)(26)$    &$0.293(38)(15)$     &$-$&$-$&$-$&-&$-0.49$ \\
Belle~\cite{Sato:2016svk}      &$-$    &$0.302(30)(11)$   &$-$  &$-$&$-$&-&$-$\\
Belle~\cite{Hirose:2016wfn}       &$-$    &$0.270(35)(_{-0.025}^{+0.028})$    &$-0.38(51)(_{-0.16}^{+0.21})$&$-$&$-$&-&$0.33$ \\
Belle~\cite{Abdesselam:2019dgh}   &$0.307(37)(16)$    &$0.283(18)(14)$    &$-$ &$-$&$-$&-&$-0.54$\\
LHCb~\cite{Aaij:2015yra}       &$-$    &$0.336(27)(30)$   &$-$  &$-$&$-$&-&$-$\\
LHCb~\cite{Aaij:2017uff,Aaij:2017deq}       &$-$    &$0.291(19)(26)(13)$   &$-$  &$-$&$-$&-&$-$\\
LHCb~\cite{Aaij:2017tyk}       &$-$    &$-$   &$-$  &$-$&$0.71(17)(18)$&-&$-$\\
Belle~\cite{Adamczyk:2019wyt,Abdesselam:2019wbt}&$-$&$-$&$-$& $0.60(8)(4)$&$-$& &$-$\\
LHCb~\cite{LHCb:2022piu} &$-$&$-$&$-$&$-$&$-$&0.242(26)(40)(59)&$-$\\
\hline
\hline
\end{tabular}
\end{scriptsize}
\label{tab:exdata}
\end{table*}

\begin{table*}
\centering
\caption{Best-fit values of the Wilson coefficients and leptoquark couplings in different new physics scenarios (without $R(\Lambda_c)$).}
\begin{scriptsize}
\begin{tabular}{ccccc}
\hline
\hline
new physics types  & Value& $\chi^2/d.o.f.$   &Correlation & Favoured?\\
\hline
$V_1$       &$(1+Re[C_{V_{1}}])^2+(Im[C_{V_{1}}])^2=1.236(38) $   &$13.55/11$  & $-$&$ \checkmark $\\
$V_2$       &$C_{V_{2}}=-0.030(34)\pm0.460(52)i $   &$12.77/11$ & $\pm0.59$&$ \checkmark $ \\
$S_1$       &$C_{S_{1}}=0.245+0.000i$   &$32.63/11$ & $-$&$\times $ \\
$S_{2}$       &$C_{S_{2}}=-0.168\pm0.744i $   &$30.10/11$  & $-$&$ \times $\\
$T$       &$C_{T}=0.014(62)\pm0.167(58)i $   &$16.61/11$  & $\pm0.98$&$ \checkmark $\\
LQ $R_2$&$(Re[y_L^{c\tau}(y_R^{b\tau})^*],Im[y_L^{c\tau}(y_R^{b\tau})^*])=(-0.811(234),\pm1.439(128))$ &12.74/11& $\pm0.82$&$\checkmark $  \\
		\multirow{2}{*}{LQ $S_1$}&$(y_L^{b\tau}(Vy_L^*)^{c\tau}),y_L^{b\tau}(y_R^{c\tau})^*)=(0.938(270),0.481(511))$  &\multirow{2}{*}{$12.52/11$} &\multirow{2}{*}{0.92}&\multirow{2}{*}{$\checkmark $}\\
		&$(y_L^{b\tau}(Vy_L^*)^{c\tau}),y_L^{b\tau}(y_R^{c\tau})^*)=(-13.227(270),-0.481(511))$ & & &\\
		\multirow{2}{*}{LQ $U_1$}&$((Vx_L)^{c\tau}(x_L^{b\tau})^*,(Vx_L)^{c\tau}(x_R^{b\tau})^*)=(0.392(85),0.061(86))$  &\multirow{2}{*}{$13.01/11$} &\multirow{2}{*}{0.78}&\multirow{2}{*}{$\checkmark $}\\
		&$((Vx_L)^{c\tau}(x_L^{b\tau})^*,(Vx_L)^{c\tau}(x_R^{b\tau})^*)=(-6.536(85),-0.061(86))$ & & &\\
		\hline
		\hline
	\end{tabular}
\end{scriptsize}
	\label{tab:coef1}
\end{table*}
The experimental data we use are listed in Table~\ref{tab:exdata}, which includes the experimental measurements of  $R(D)$, $R(D^*)$, $P_\tau(D^*)$, $F_L^{D^{*}}$, $R(J/\psi)$ and $R(\Lambda_c)$ together with the correlations. In the fit, we also consider the theoretical correlations of these observables as in \cite{Cheung:2020sbq}. The results of the fits (imposing $\mathcal B(B_c\to J/\psi)<30\%$) are listed in Table~\ref{tab:coef1} (without $R(\Lambda_c)$) and \ref{tab:coef2} (with $R(\Lambda_c)$). {We use the theoretical formulas for $R(\Lambda_c)$ presented in \cite{Datta:2017aue} based on the lattice QCD calculation of form factors~\cite{Detmold:2015aaa}. Allowed parameter space at the C.L. of $1\sigma$ and $2\sigma$ for the Wilson coefficients constrained by  the measurements of $R(D^{(*)})$, $R(J/\psi)$, $R(\Lambda_c)$ and other $b\to c\tau\nu$ observables are shown in Figure~\ref{fig:2sigmawil} for model independent scenarios and Figure~\ref{fig:2sigmalep} for the leptoquark models. The allowed (forbidden) regions by $\mathcal B(B_c \to \tau\nu)<30\%$ are in light (dark) grey.  Using the results in Table~\ref{tab:coef2}, we also give predictions for all the observables, which are shown in the next section.
\begin{table*}
\centering
\caption{Best-fit values of the Wilson coefficients and leptoquark couplings in different new physics scenarios (with $R(\Lambda_c)$).}
\begin{scriptsize}
\begin{tabular}{ccccc}
\hline
\hline
new physics types  & Value& $\chi^2/d.o.f.$   &Correlation&Favoured?\\
\hline
$V_1$       &$(1+Re[C_{V_{1}}])^2+(Im[C_{V_{1}}])^2=1.222(38) $   &$18.39/12$  & $-$&$ \checkmark $\\
$V_2$       &$C_{V_{2}}=-0.031(34)\pm0.445(53)i $   &$17.59/12$ & $\pm0.59$&$ \checkmark $ \\
$S_1$       &$C_{S_{1}}=0.230+0.000i$   &$36.08/12$ & $-$&$\times $ \\
$S_{2}$       &$C_{S_{2}}=-0.183\pm0.735i $   &$33.68/12$  & $-$&$ \times $\\
$T$       &$C_{T}=0.000(62)\pm0.148(71)i $   &$21.69/12$  & $\pm0.98$&$ \checkmark $\\
LQ $R_2$    &$(Re[y_L^{c\tau}(y_R^{b\tau})^*],Im[y_L^{c\tau}(y_R^{b\tau})^*])=(-0.807(234),\pm1.436(128))$ &15.56/12& $\pm0.81$&$\checkmark $  \\
\multirow{2}{*}{LQ $S_1$}&$(y_L^{b\tau}(Vy_L^*)^{c\tau}),y_L^{b\tau}(y_R^{c\tau})^*)=(0.948(269),0.577(523))$  &\multirow{2}{*}{$16.93/12$} &\multirow{2}{*}{0.92}&\multirow{2}{*}{$\checkmark $}\\
		&$(y_L^{b\tau}(Vy_L^*)^{c\tau}),y_L^{b\tau}(y_R^{c\tau})^*)=(-13.236(269),-0.577(523))$ & & &\\
\multirow{2}{*}{LQ $U_1$}&$((Vx_L)^{c\tau}(x_L^{b\tau})^*,(Vx_L)^{c\tau}(x_R^{b\tau})^*)=(0.377(84),0.067(86))$  &\multirow{2}{*}{$17.74/12$} &\multirow{2}{*}{0.78}&\multirow{2}{*}{$\checkmark $}\\
		&$((Vx_L)^{c\tau}(x_L^{b\tau})^*,(Vx_L)^{c\tau}(x_R^{b\tau})^*)=(-6.521(84),-0.067(86))$ & & &\\
		\hline
		\hline
	\end{tabular}
\end{scriptsize}
	\label{tab:coef2}
\end{table*}
By comparing the fit results with those in \cite{Cheung:2020sbq} where $B_c\to J/\psi$ form factors calculated in the covariant light-front quark model are used~\cite{Wang:2008xt}, we find that the change of the fit results are tiny, because the $B_c\to J/\psi$ form factors are only relevant to one experimental measurement, i.e. $R(J/\psi)$. In fact, all the $\chi^{2}$ values have been slightly reduced, which suggests that the lattice results are slightly more compatible with the measurements. Due to the tiny impact of $B_c\to J/\psi$ form factors on the global analysis, from Table~\ref{tab:coef1} it is clear that without considering $R(\Lambda_c)$ the conclusions in \cite{Cheung:2020sbq} on the new physics scenarios still hold: the $S_{1}$ scenario is ruled out by the $b\to c\tau\nu$ data, and the $ S_{2} $ scenario is not supported by the constraint from the pure leptonic $B_c$ decay ($\mathcal B(B_c\to\tau\nu)<30\%$). Regarding the leptoquark models, we find that the previously disfavoured $R_2$ model, can now be favoured due to the less stringent constraint from $B_c\to\tau\nu$ decay.
\begin{figure}
\begin{center}
\includegraphics[scale=0.16]{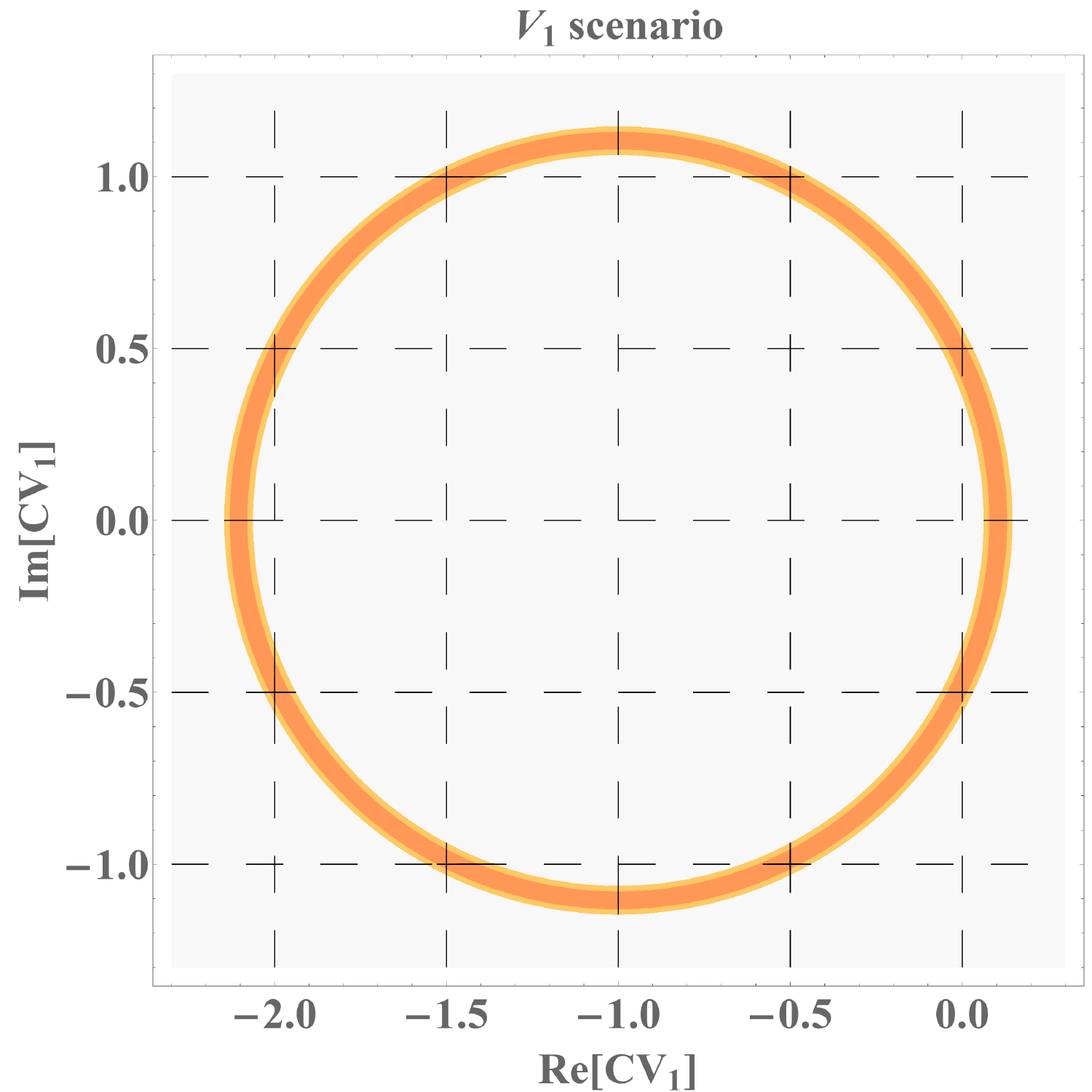}
\includegraphics[scale=0.16]{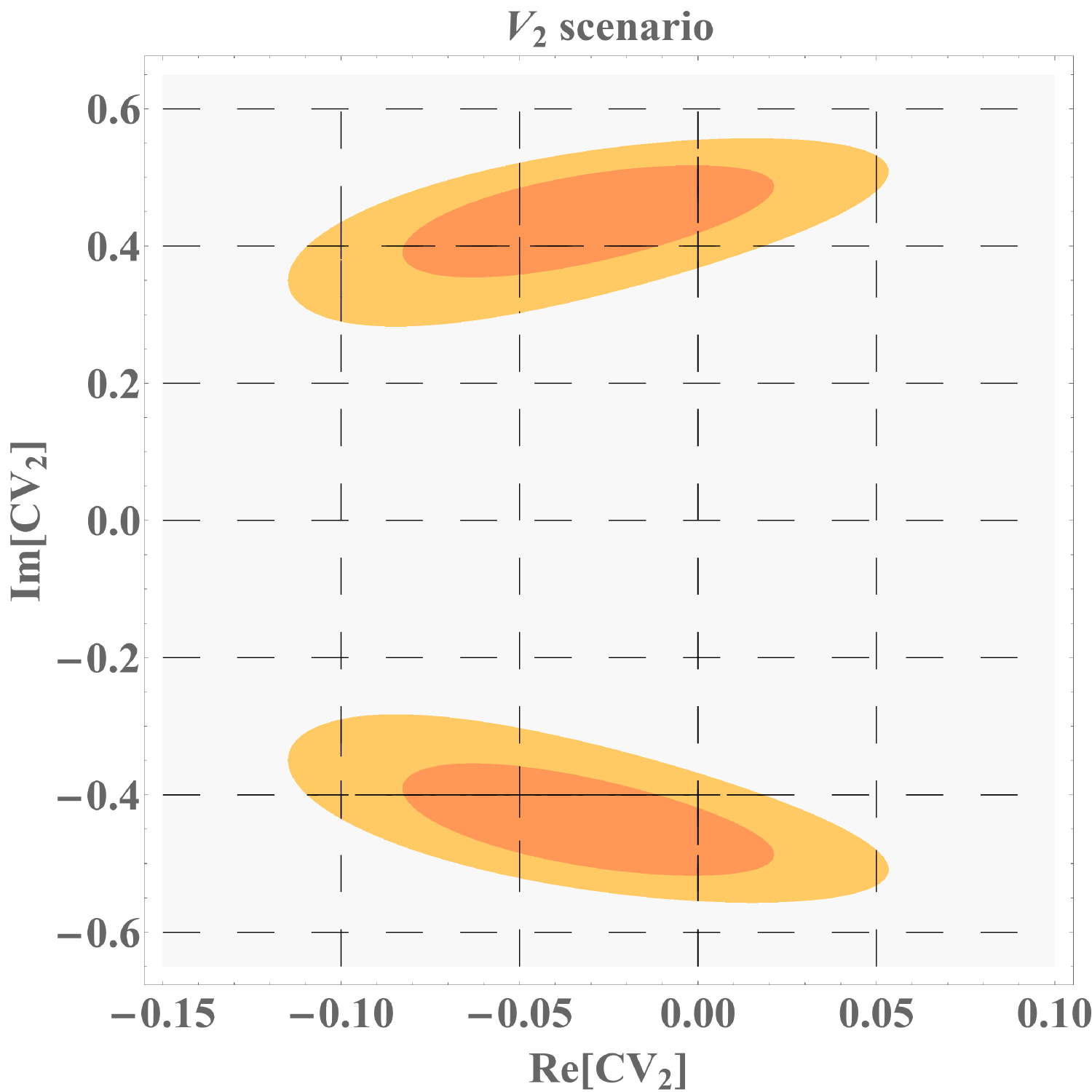}
\includegraphics[scale=0.16]{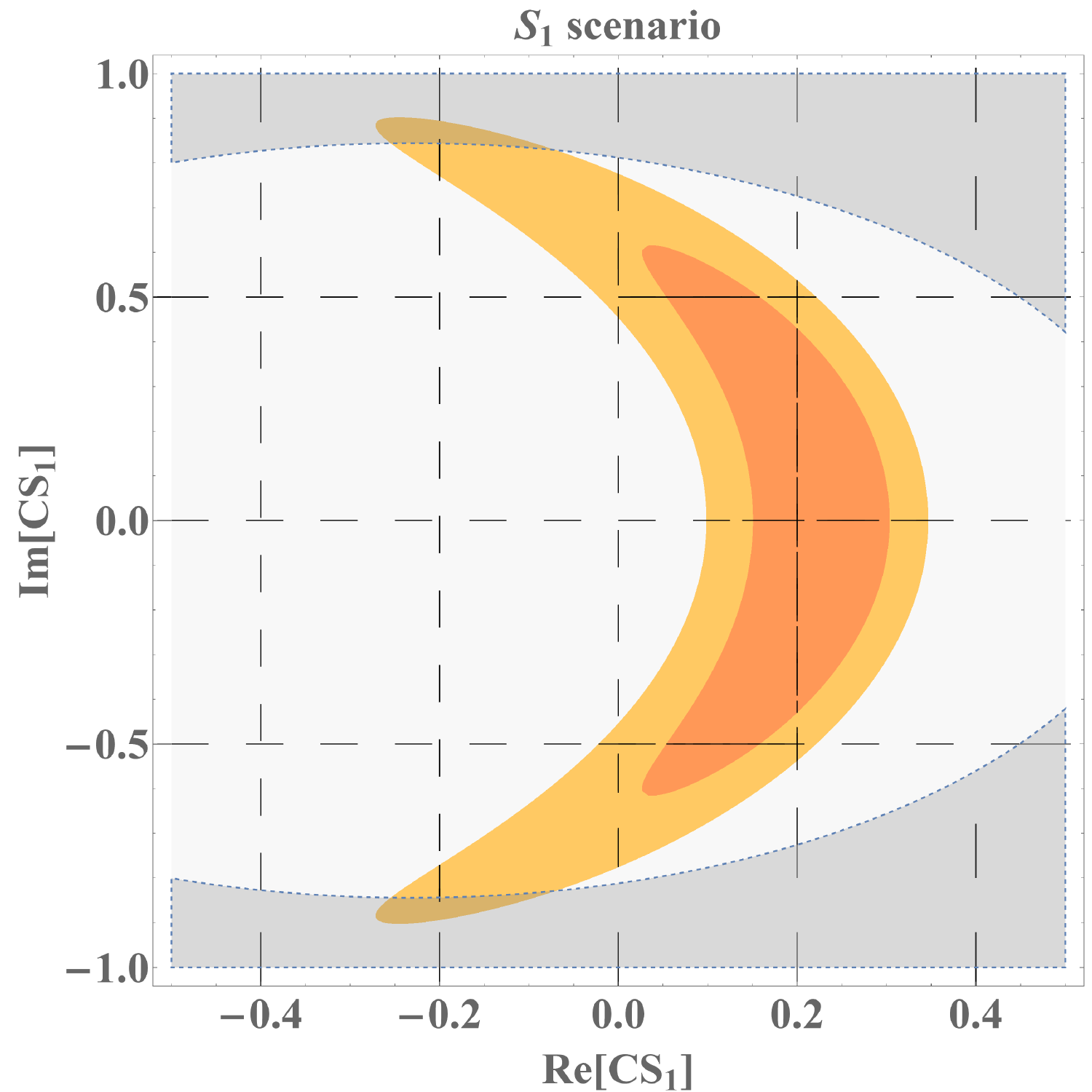}
\includegraphics[scale=0.16]{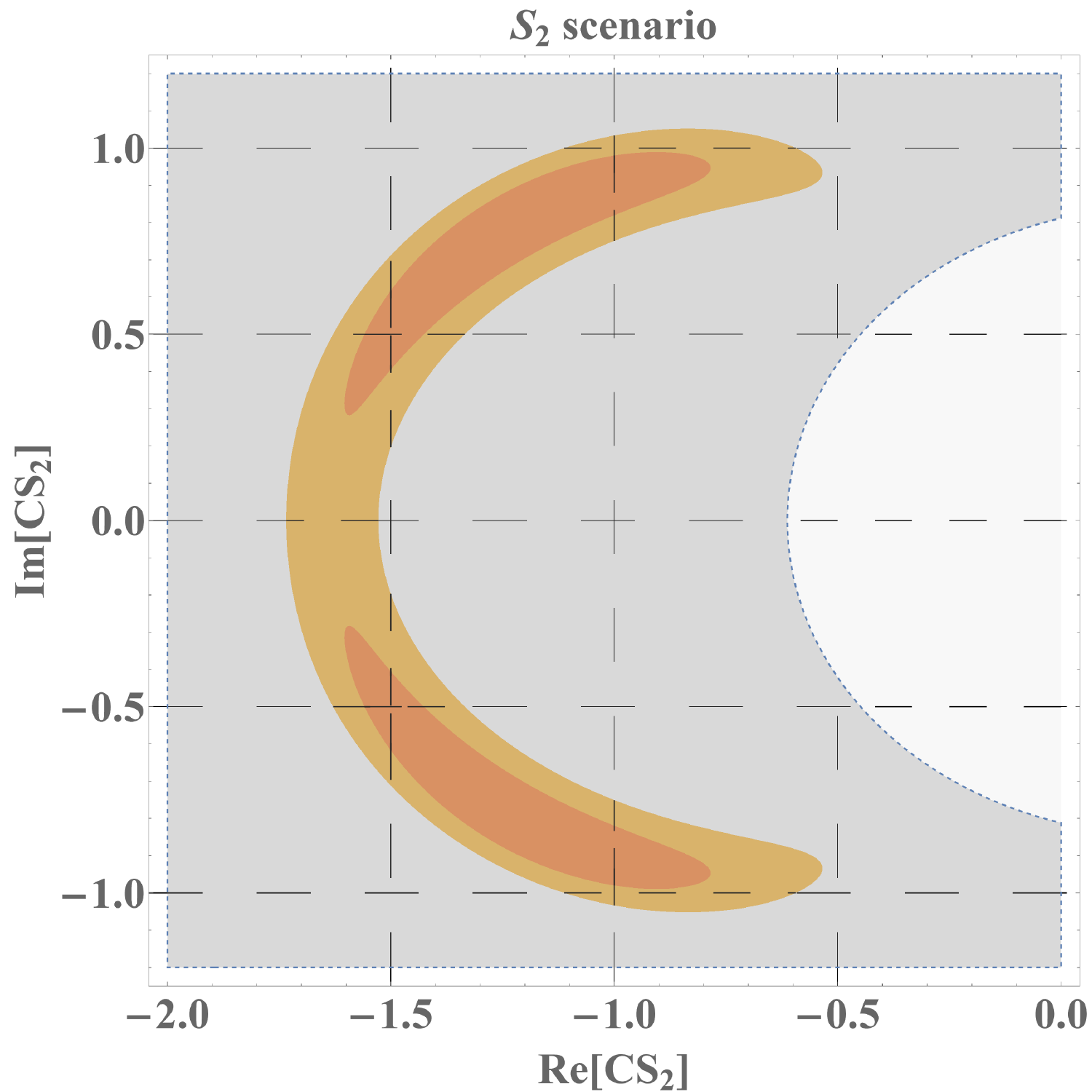}
\includegraphics[scale=0.16]{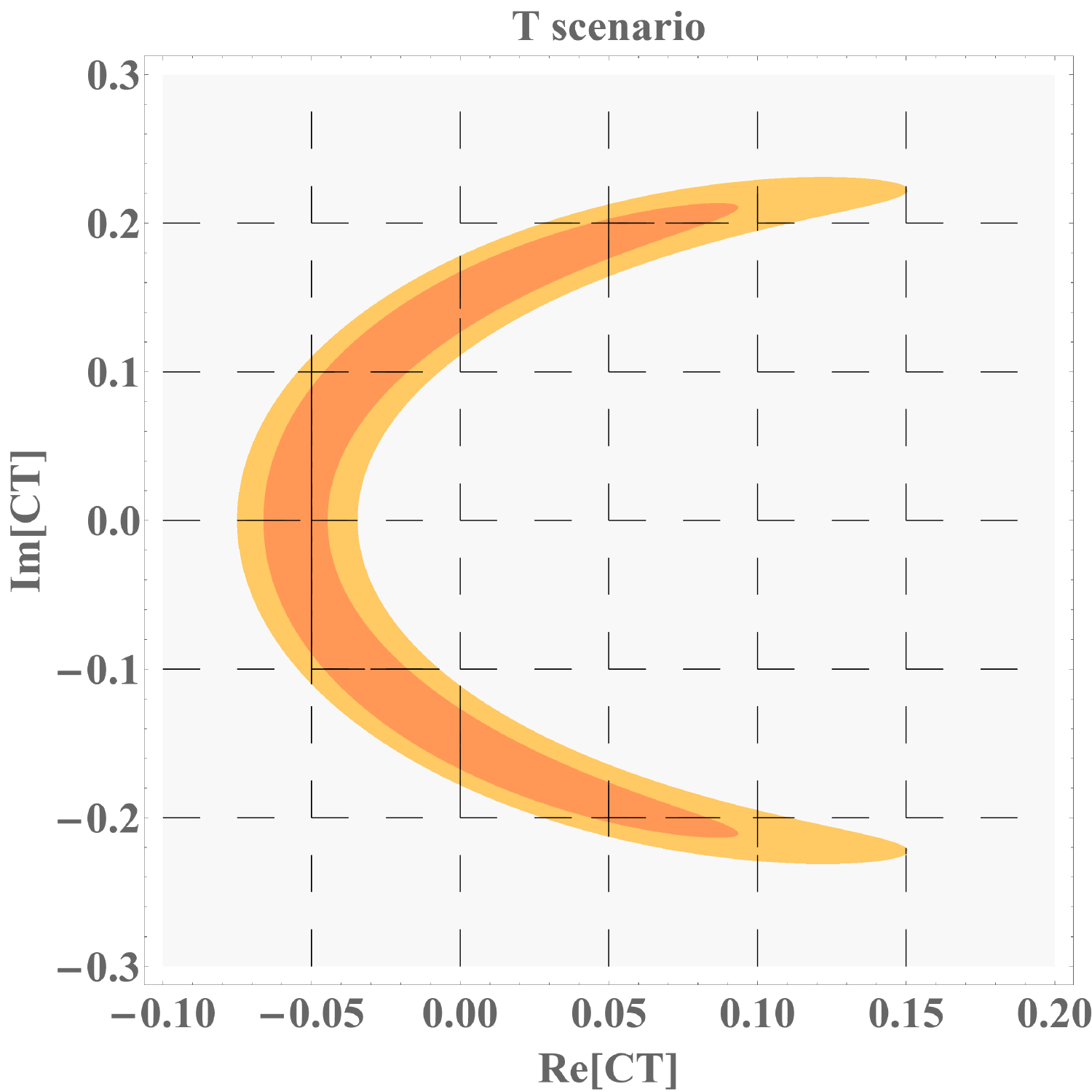}
\caption{Allowed parameter space at the C.L. of $1\sigma$ and $2\sigma$ for the Wilson coefficients by the measurements of $R(D^{(*)})$, $R(J/\psi)$, $R(\Lambda_c)$ and other $b\to c\tau\nu$ observables. The allowed (forbidden) regions by $\mathcal B(B_c \to \tau\nu)<30\%$ are in light (dark) grey.
}
\label{fig:2sigmawil}
\end{center}
\end{figure}

\begin{figure}
	\begin{center}
		\includegraphics[scale=0.16]{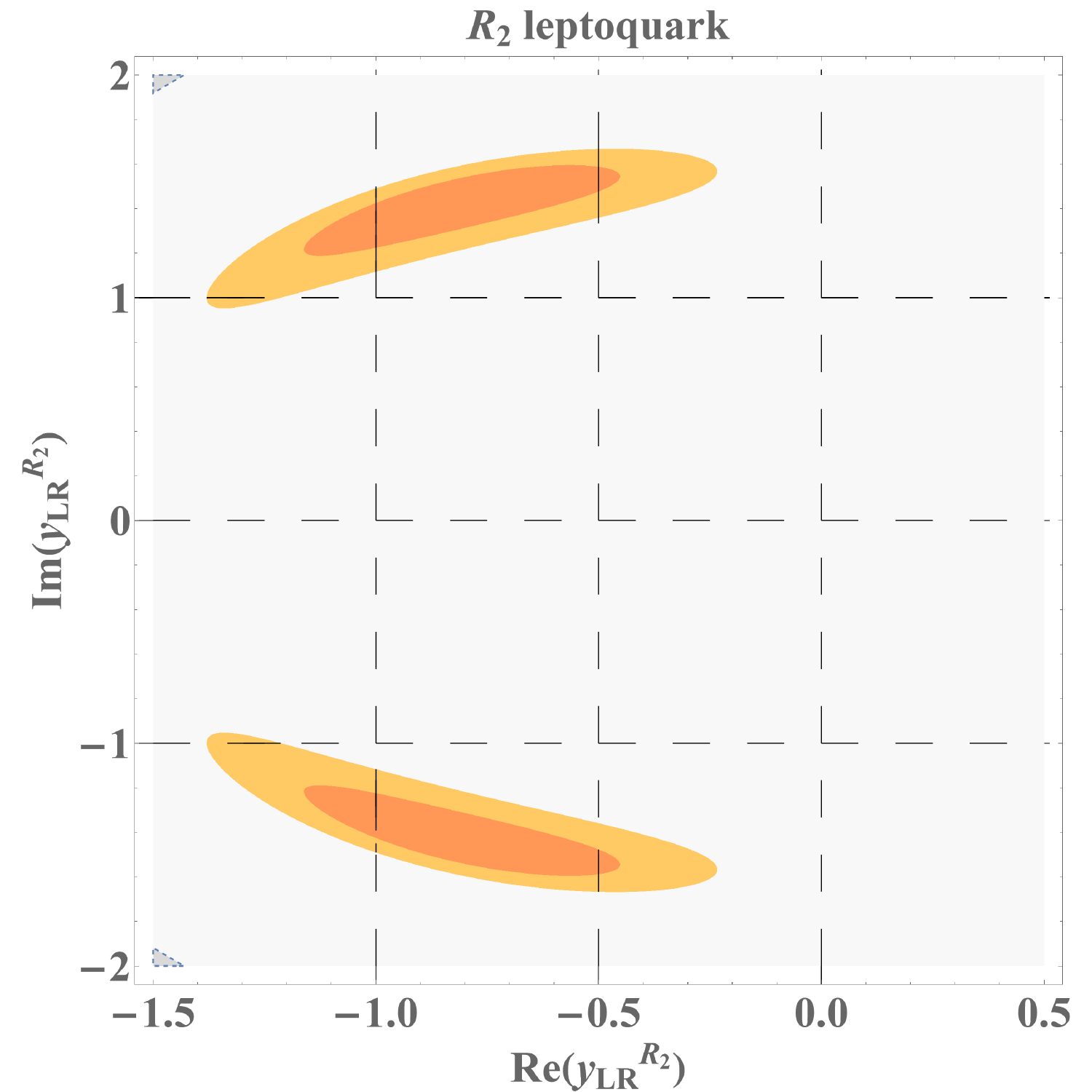}
		\includegraphics[scale=0.16]{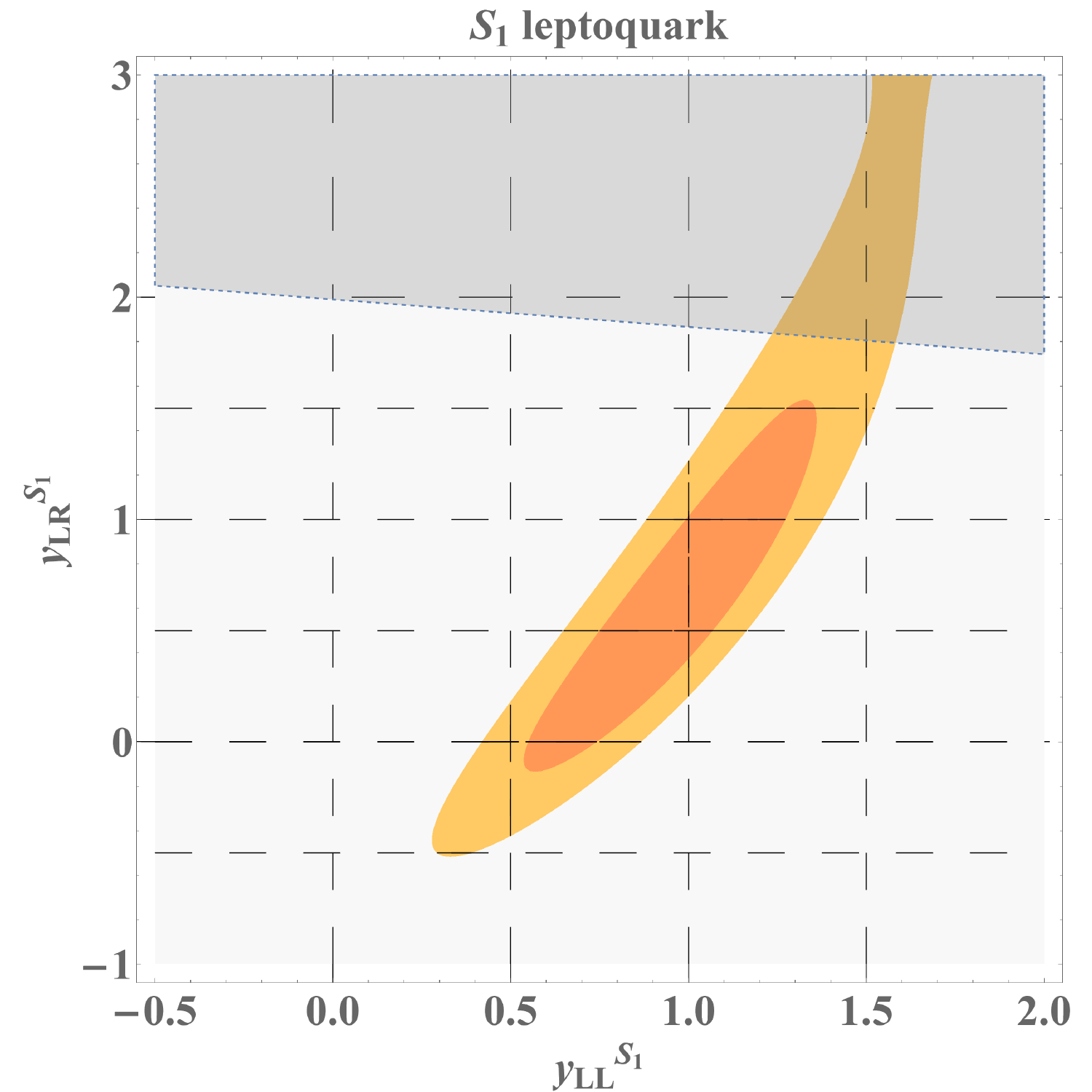}
		\includegraphics[scale=0.16]{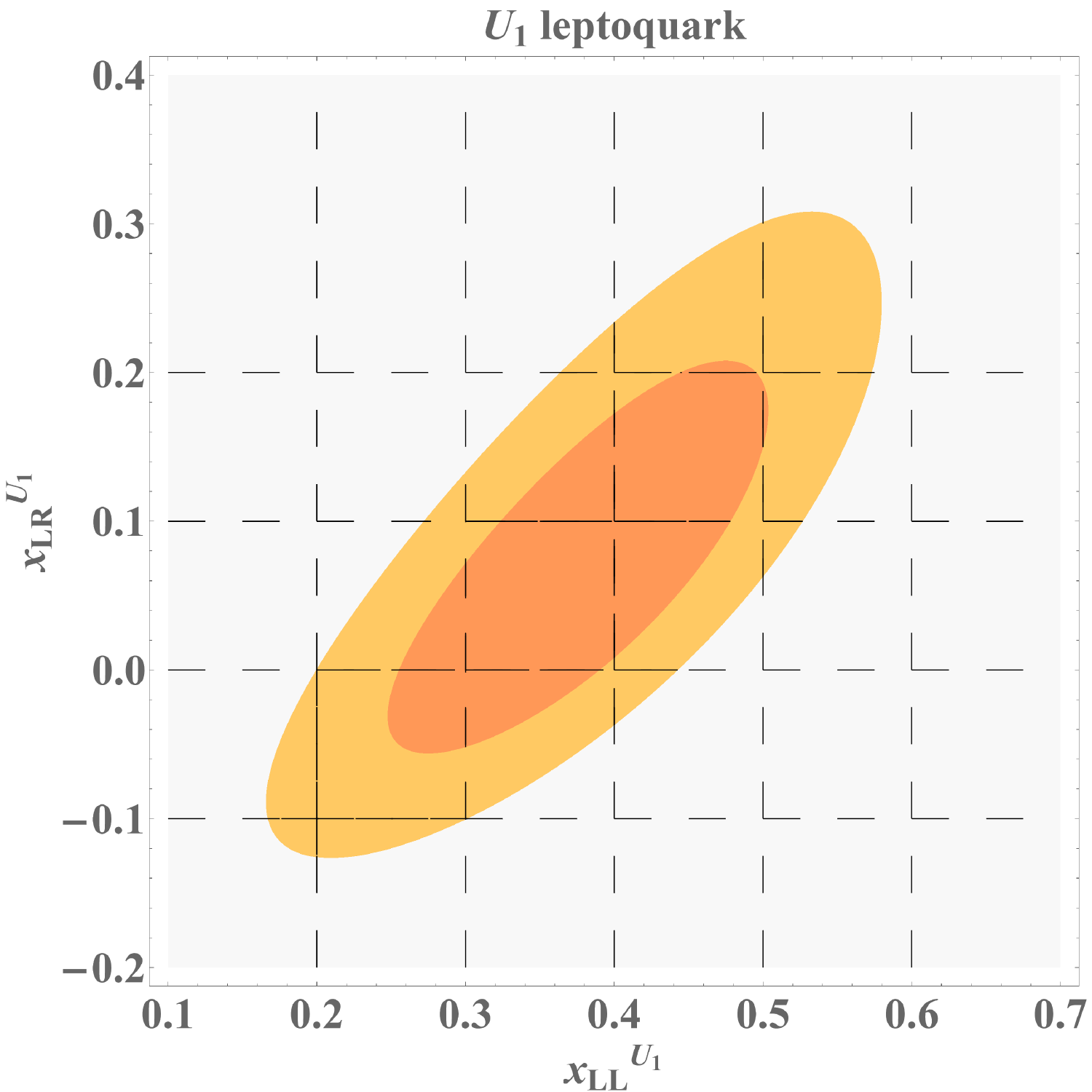}
		\caption{Allowed parameter space at the C.L. of $1\sigma$ and $2\sigma$ for the leptoquark couplings by the measurements of $R(D^{(*)})$, $R(J/\psi)$, $R(\Lambda_c)$ and other $b\to c\tau\nu$ observables. The allowed (forbidden) regions by $\mathcal B(B_c \to \tau\nu)<30\%$ are in light (dark) grey.}\label{fig:2sigmalep}
	\end{center}
\end{figure}
Adding $R(\Lambda_c)$, as shown in Table~\ref{tab:coef2} and Figure~\ref{fig:2sigmawil} and \ref{fig:2sigmalep}, the above conclusions all remain unchanged, although $\chi^2/d.o.f.$ increases for each scenario/model as a result of the fact that the central value of $R(\Lambda_c)$ shows a different tendency in the deviation from the SM compared to $R(D^{(*)})$ and $R(J/\psi)$. Under the constraint $\mathcal B(B_c\to\tau\nu)<30\%$, the $R_2$ leptoquark model is favoured and interestingly it gives the smallest $\chi^2/d.o.f.$ among all scenarios/models after taking $R(\Lambda_c)$ into account.
\section{predictions for the observables}\label{sec:PRE}
With the fitted results in Section~\ref{sec:CWC}, we make predictions for some observables that can be used to probe the new physics effects. We not only study the LFU ratio $R(J/\psi)$, but also several other polarization and angular observables, including the longitudinal polarization fraction of the $\tau$ lepton $P_{\tau}(J/\psi)$, the longitudinal polarization fraction of the $J/\psi$ meson $F_L(J/\psi)$ and the forward-backward asymmetry of the $\tau$ lepton $\mathcal{A}_{FB}(J/\psi)$. These observables are defined as~\footnote{Note that there is a sign difference in the definitions of the $\tau$ polarization fraction between our work and \cite{Harrison:2020nrv}.}
\begin{eqnarray}
            \label{eq:Ptau}
 	        P_\tau(J/\psi) &=&  \frac{\Gamma(\lambda_\tau=1/2)-\Gamma(\lambda_\tau=-1/2)}{\Gamma(\lambda_\tau=1/2)+\Gamma(\lambda_\tau=-1/2)} \,,\\
            \label{eq:PJpsi}
 	        F_L(J/\psi) &=&\frac{\Gamma(\lambda_{J/\psi}=0)}{\Gamma(\lambda_{J/\psi}=0)+\Gamma(\lambda_{J/\psi}=1)+\Gamma(\lambda_{J/\psi}=-1)}\,,\\
            \label{eq:AFB}
 	        \A_{\rm FB}(J/\psi) &=&\left(  \int_0^1 \frac{d\Gamma} { d\cos\th_{\ell}}d\cos\th_{\ell}-\int^0_{-1}\frac{d\Gamma}{ d\cos\th_{\ell}}d\cos\th_{\ell}\right) \bigg/ \Gamma\,,
          \end{eqnarray}
          where $\Gamma$ is the decay width of $B_c\to J/\psi\tau\nu$, $\lambda_\tau$ is the $\tau$ helicity in the rest frame of the leptonic system, $\lambda_{J/\psi}$ is the helicity of $J/\psi$ in the $B_{c}$ rest frame, and $\theta_{\ell}$ is the angle between the momentum of $\tau$ and $B_{c}$ in the rest frame of $\tau\nu$.
\begin{eqnarray}
            \label{eq:RJpsiq2}
             R_{J/\psi}(q^2)&=&\frac{d\mathcal B(B_{c}\to J/\psi\tau\nu)/dq^2}{d\mathcal B(B_{c}\to J/\psi\mu\nu)/dq^2}\,,\\
            \label{eq:Ptauq2}
            P^{J/\psi}_\tau(q^{2}) &=& { \left(\frac{d\Gamma(\lambda_\tau=1/2)}{dq^{2}} - \frac{d\Gamma(\lambda_\tau=-1/2)}{dq^{2}}\right) \bigg/ \left(\frac{d\Gamma(\lambda_\tau=1/2) }{dq^{2}}+ \frac{d\Gamma(\lambda_\tau=-1/2)}{dq^{2}}\right) } \,,\nonumber\\ \\
            \label{eq:PJpsiq2}
            F_L^{J/\psi}(q^{2}) &=& { \frac{d\Gamma(\lambda_{J/\psi}=0)}{dq^{2}} \bigg/ \left(\frac{d\Gamma(\lambda_{J/\psi}=0)}{dq^{2}} + \frac{d\Gamma(\lambda_{J/\psi}=1)}{dq^{2}} + \frac{d\Gamma(\lambda_{J/\psi}=-1)}{dq^{2}}\right) } \,,\\
            \A^{J/\psi}_{\rm FB}(q^{2}) &=&\left(  \int_0^1 \frac{d\Gamma} { dq^{2} d\cos\th_{\ell}}d\cos\th_{\ell}-\int^0_{-1}\frac{d\Gamma}{ dq^{2} d\cos\th_{\ell}}d\cos\th_{\ell}\right) \bigg/ \frac{d\Gamma}{dq^{2} }\,  .
            \label{eq:AFBq2}
            \end{eqnarray}
Taking the fitted Wilson coefficients from Table~\ref{tab:coef2} for the new physics scenarios/models we compute the observables in Equation~(\ref{eq:Rjpsi}) and (\ref{eq:Ptau})-(\ref{eq:AFBq2}). We exclude the $S_1$ and $S_2$ scenarios in our analyses due to their large $\chi^2$, and focus on the $V_1$, $V_2$ and $T$ model independent scenarios. Since the $R_{2}$ leptoquark is also favoured by the fit imposing $\mathcal B(B_c\to\tau\nu)<30\%$, we give predictions for all three leptoquark models $R_{2}$, $S_{1}$ and $U_{1}$. The results for the integrated observables are listed in Table~\ref{tab:obs}. The SM predictions for these observables are given in \cite{Harrison:2020nrv}, which we confirm and list in the first row. From Table~\ref{tab:obs} we find that the predictions for $R(J/\psi)$ in all scenarios/models are larger than the SM prediction because the experimental measurements of the LFU ratios (except $R(\Lambda_c)$) exceed the SM predictions. Among the other observables, we find $P_\tau(J/\psi)$ in $R_2$ leptoquark model, and $\A_{FB}(J/\psi)$ in $V_2$ scenario and $R_2$ leptoquark model are distinguishable from the SM predictions. Moreover, in $T$ scenario, all observables are distinct from the SM predictions in their central values, but they also associate with large errors, which reduce the distinguishability of new physics.
\begin{table*}
	\centering
	\caption{Predictions for $B_{c}\to J/\psi\tau\nu$ observables. The first and the second errors are respectively due to the form factors and the Wilson coefficients.}
\begin{small}
	\begin{tabular}{c c c c c}
		\hline
		\hline
		& $R(J/\psi)$ & $P_\tau(J/\psi)$ & $F_L(J/\psi)$ & $\A_{FB}(J/\psi)$ \\
		\hline
           SM  &$0.258(4)$&$-0.518(7)$&$0.442(9)$&$-0.058(12)$\\
		$V_{1}$&0.316(5)(10)&$-0.518(7)(0)$&0.442(9)(0)&$-0.058(12)(0)$\\
		$V_{2}$&0.325(5)(14)&$-0.518(7)(0)$&0.444(9)(2)&$0.007(9)(9)$\\
	    $T$&0.340(9)(50)&$-0.382(10)(81)$&0.384(8)(36)&$0.014(9)(24)$\\
        LQ $R_2$ &0.314(5)(22)&-0.438(9)(23)&0.439(9)(1)&0.007(11)(20)\\
        LQ $S_1$ &0.322(4)(11)&-0.506(10)(16)&0.457(10)(14)&-0.066(13)(7)\\
        LQ $U_1$ &0.324(5)(14)&-0.529(7)(14)&0.438(9)(5)&-0.064(12)(8)\\
		\hline
		\hline
	\end{tabular}
\end{small}
	\label{tab:obs}
\end{table*}
Our results for the $q^2$ dependent observables $R_{J/\psi}(q^2)$, $P_\tau^{J/\psi}(q^2)$, $F_L^{J/\psi}(q^2)$ and $\mathcal A_{FB}^{J/\psi}(q^2)$ are plotted in Figure~\ref{fig:rBcq2}-\ref{fig:AFB}. In all these figures, we show the predictions of the SM with the red solid line with uncertainties from form factors, and predictions of new physics with the blue dashed lines, which include uncertainties from both Wilson coefficients and form factors.

From Figure~\ref{fig:rBcq2}, we can see that in all these three new physics scenarios $R_{J/\psi}(q^2)$ is larger than that in the SM in full $q^2$ region, therefore the $q^2$ distribution of $R_{J/\psi}$ is very useful for testing new physics, although not as helpful to separate different new physics scenarios. Moreover, from Figure~\ref{fig:Ptauq2}-\ref{fig:AFB}, it shows that $\O_{V_1}$ has no effects on $P_\tau^{J/\psi}(q^2)$, $F_L^{J/\psi}(q^2)$ and $\mathcal A_{FB}^{J/\psi}(q^2)$ because its contributions to these observables get cancelled. Due to tiny new physics effects as well as the uncertainties from form factors and experiments, the $V_2$ scenario is indistinguishable from the SM using $P_\tau^{J/\psi}(q^2)$ and $F_L^{J/\psi}(q^2)$, while as shown in Figure~\ref{fig:AFB}, $\mathcal A_{FB}^{J/\psi}(q^2)$ in the $V_2$ scenario is larger than that in the SM. The $T$ scenario has distinctive features: $P_\tau^{J/\psi}(q^2)$ is smaller than the SM prediction at low $q^2$ but larger at high $q^2$; the prediction of $F_L^{J/\psi}(q^2)$ is smaller than the SM prediction in full $q^2$ region.
\begin{figure}
\begin{center}
\includegraphics[scale=0.39]{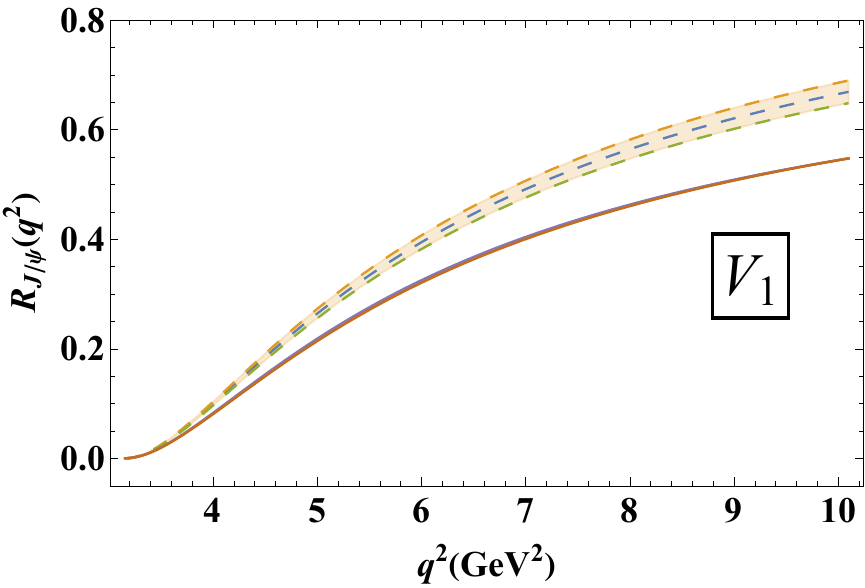}
\includegraphics[scale=0.39]{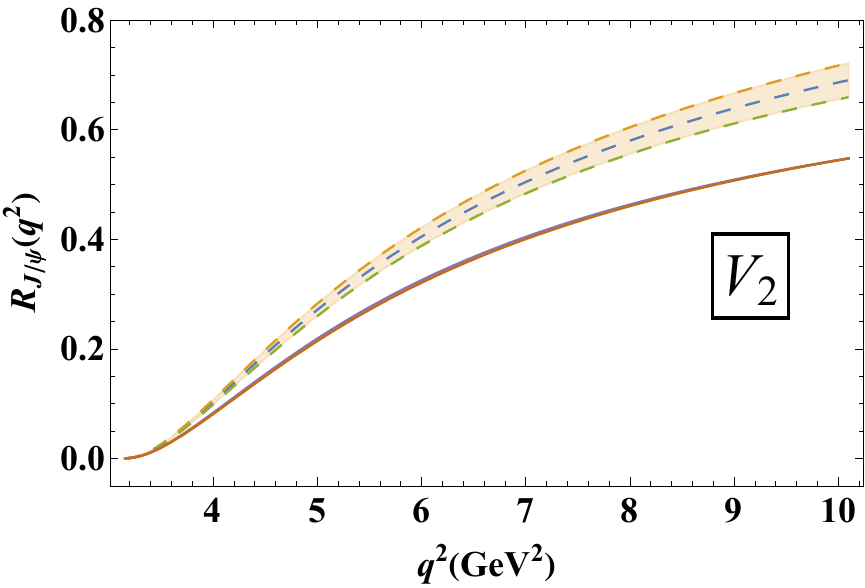}
\includegraphics[scale=0.39]{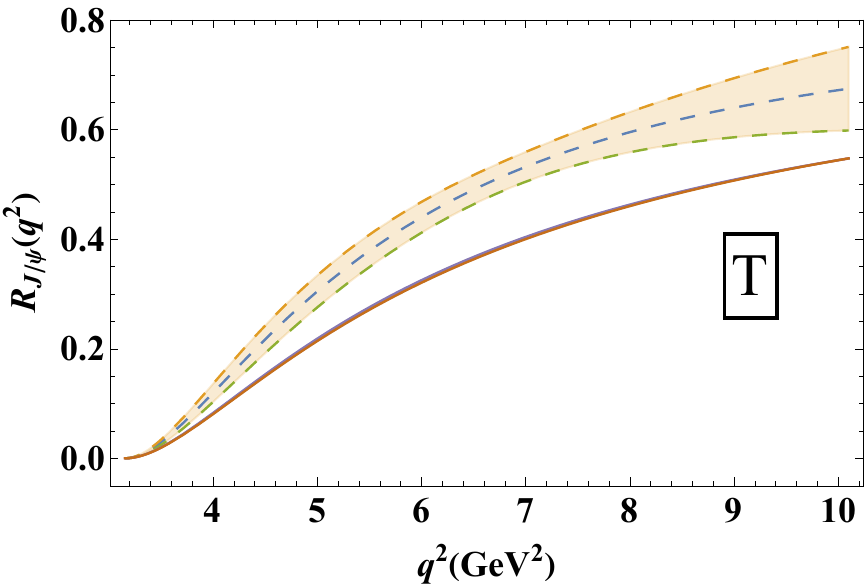}
\caption{
Predictions for the differential ratio $R_{J/\psi}(q^2)$ in the SM (red solid lines) and in the new physics scenarios (blue dashed lines)  corresponding to the best-fit Wilson coefficients. The bands include errors from form factors (for SM and new physics) and Wilson coefficients (for new physics).}
\label{fig:rBcq2}
\end{center}
\end{figure}

\begin{figure}
\begin{center}
\includegraphics[scale=0.39]{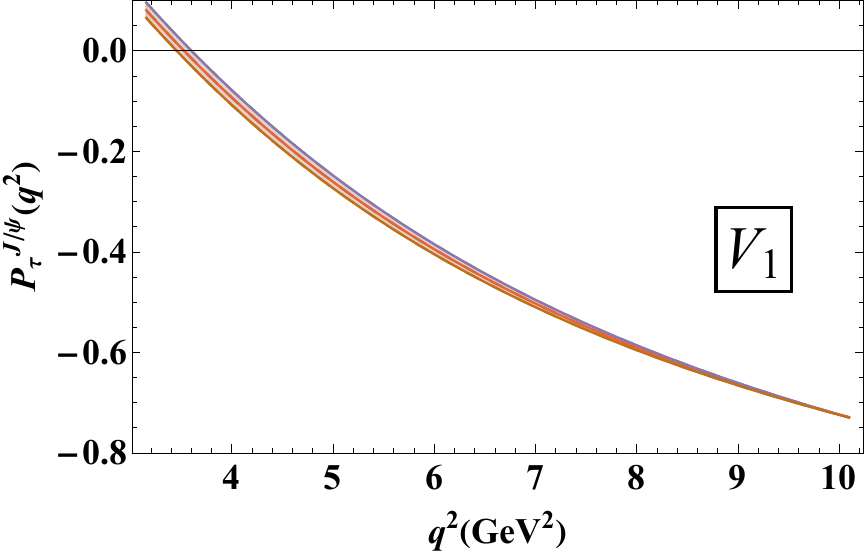}
\includegraphics[scale=0.39]{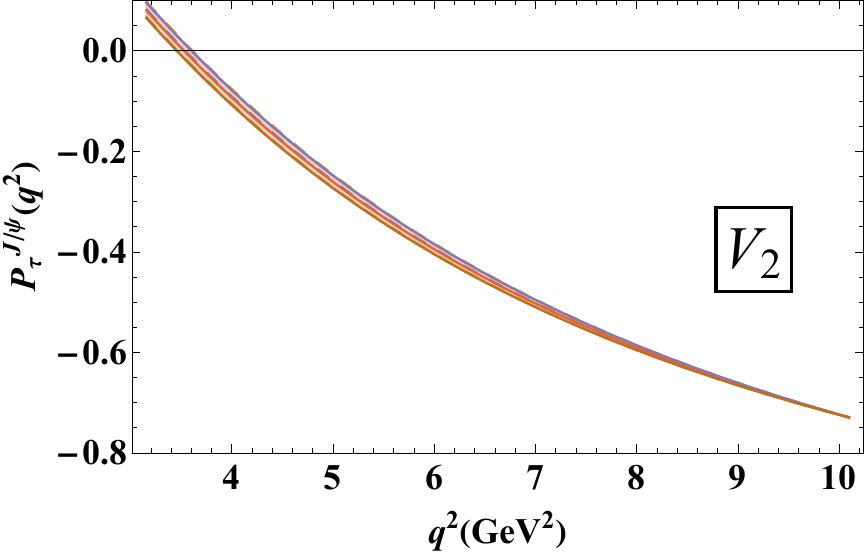}
\includegraphics[scale=0.39]{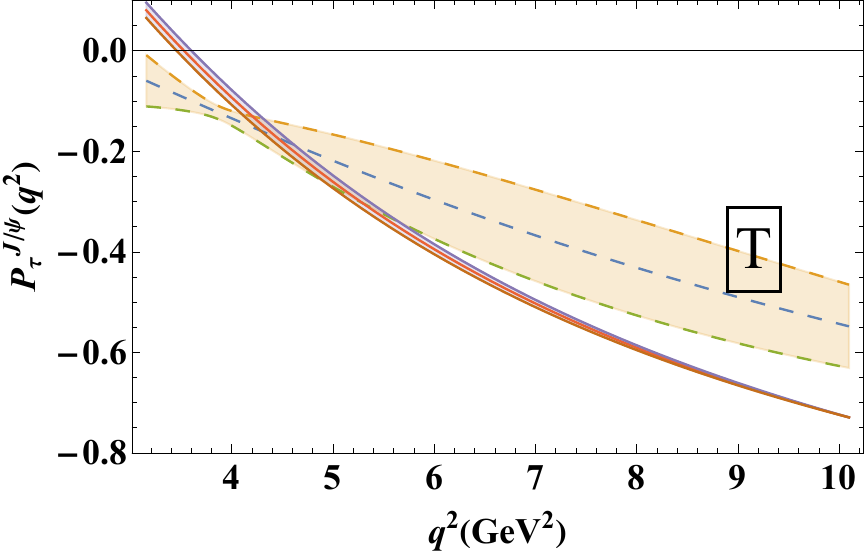}
\caption{
Predictions for the differential polarization fraction $P_\tau^{J/\psi}(q^2)$ in the SM (red solid lines) and in the new physics scenarios (blue dashed lines)  corresponding to the best-fit Wilson coefficients. The bands include errors from form factors (for SM and new physics) and Wilson coefficients (for new physics).}
\label{fig:Ptauq2}
\end{center}
\end{figure}

\begin{figure}
	\begin{center}
		\includegraphics[scale=0.39]{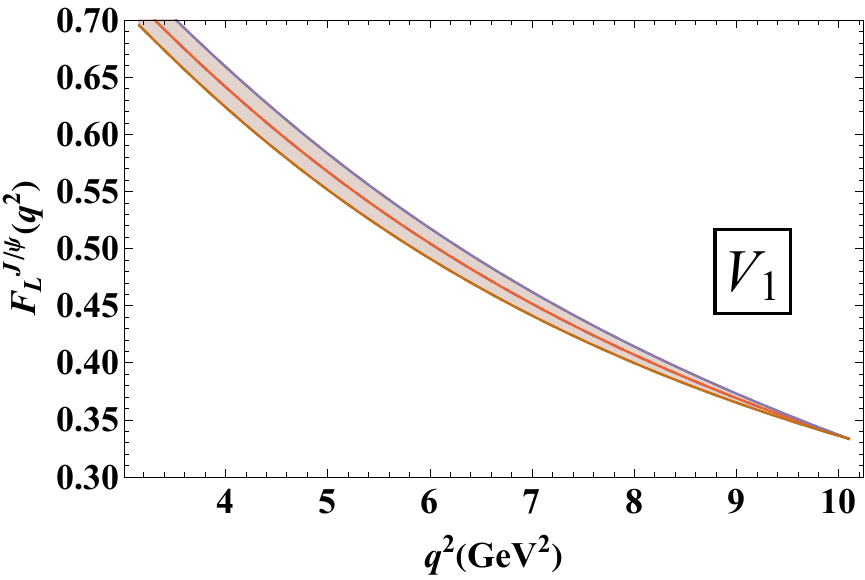}
		\includegraphics[scale=0.39]{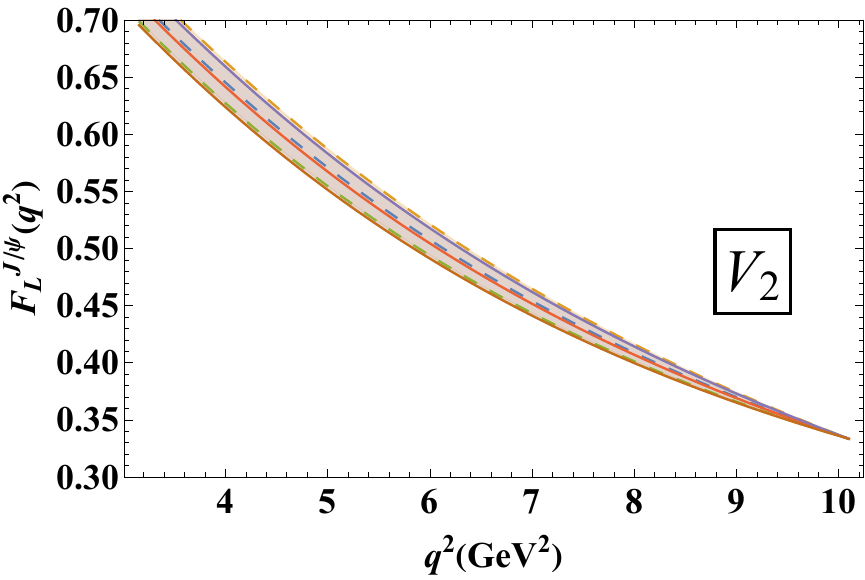}
		\includegraphics[scale=0.39]{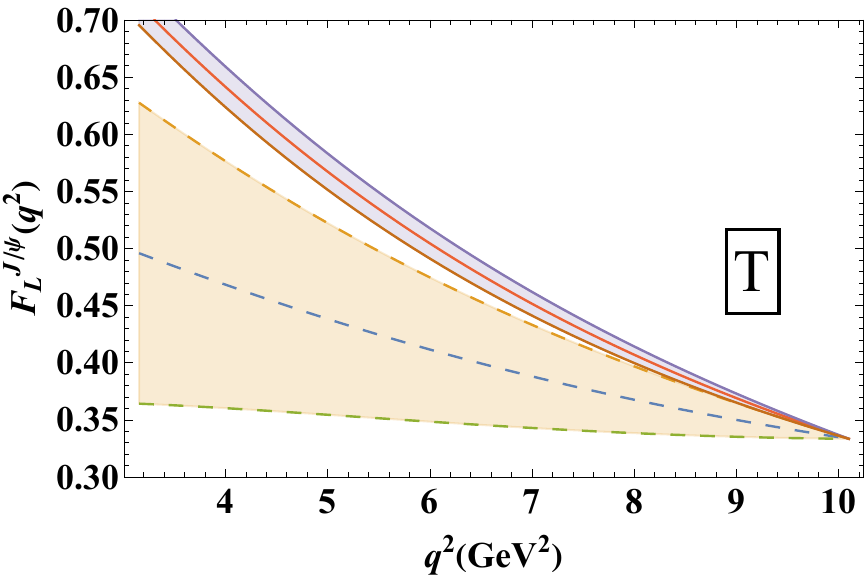}
		\caption{Predictions for the differential polarization fraction $F_L^{J/\psi}(q^2)$ in the SM (red solid lines) and in the new physics scenarios (blue dashed lines)  corresponding to the best-fit Wilson coefficients. The bands include errors from form factors (for SM and new physics) and Wilson coefficients (for new physics).}
		\label{fig:Pq2}
	\end{center}
\end{figure}
\begin{figure}
	\begin{center}
		\includegraphics[scale=0.39]{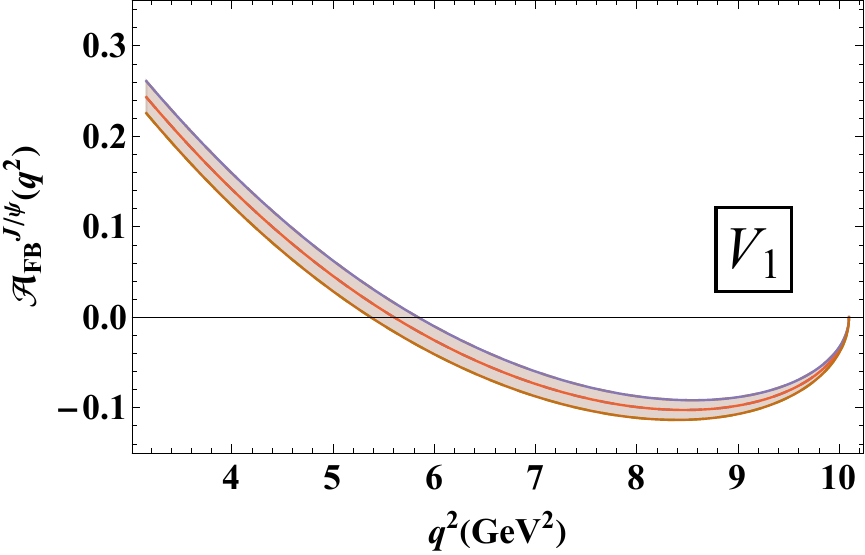}
		\includegraphics[scale=0.39]{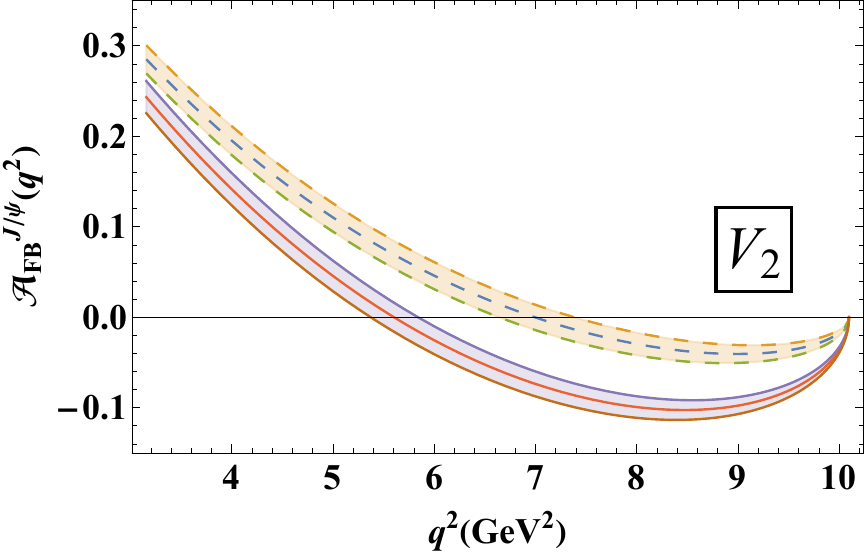}
		\includegraphics[scale=0.39]{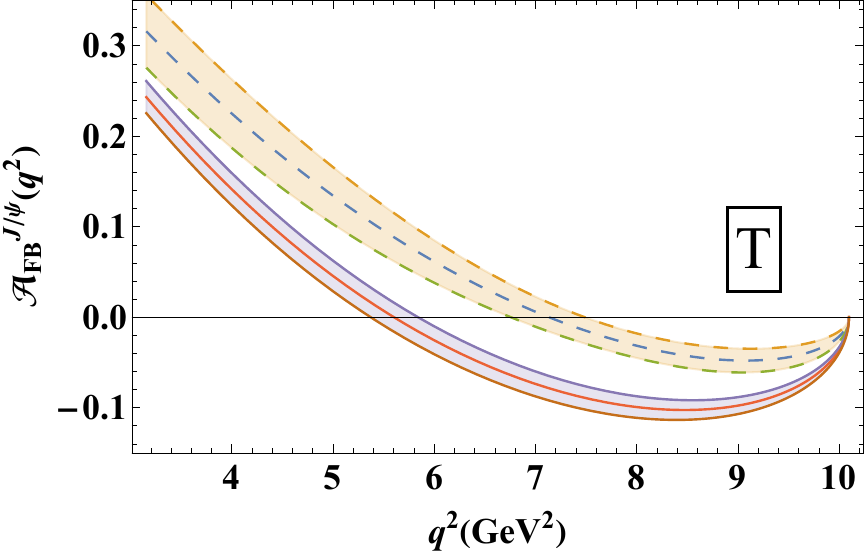}
		\caption{Predictions for the differential forward-backward asymmetry $\mathcal A_{FB}^{J/\psi}(q^2)$ in the SM (red solid lines) and in the new physics scenarios (blue dashed lines)  corresponding to the best-fit Wilson coefficients. The bands include errors from form factors (for SM and new physics) and Wilson coefficients (for new physics).}
		\label{fig:AFB}
	\end{center}
\end{figure}
\newpage
Then we discuss the leptoquark models. Our results for the $q^2$ dependent observables $R_{J/\psi}(q^2)$, $P_\tau^{J/\psi}(q^2)$, $F_L^{J/\psi}(q^2)$ and $\mathcal A_{FB}^{J/\psi}(q^2)$ are depicted in Figure~\ref{fig:lpRq2}-\ref{fig:lpAFB} for the leptoquark models. In all these figures, we show the predictions of the SM with the red solid line with uncertainties from form factors, and predictions of new physics with the blue dashed lines, which include uncertainties from both Wilson coefficients and form factors. The three types of leptoquarks give similar predictions for $R_{J/\psi}(q^2)$: the new physics predictions are larger than SM predictions especially in high $q^{2}$ region, as shown in Figure~\ref{fig:lpRq2}. In contrast, using the other three observables, $S_1$ and $U_1$ leptoquark models cannot be differentiated from the SM within the errors, but $R_2$ can be distinguished by using $P_\tau^{J/\psi}(q^2)$ and $\mathcal A_{FB}^{J/\psi}(q^2)$ in central and high $q^2$ region, as shown in Figure~\ref{fig:lpPtauq2}-\ref{fig:lpAFB}. The reason why $S_{1}$ and $U_{1}$ leptoquark models have similar phenomenological implications is that their dominant contributions are both from $\O_{V_1}$.
\begin{figure}
	\begin{center}
        \includegraphics[scale=0.39]{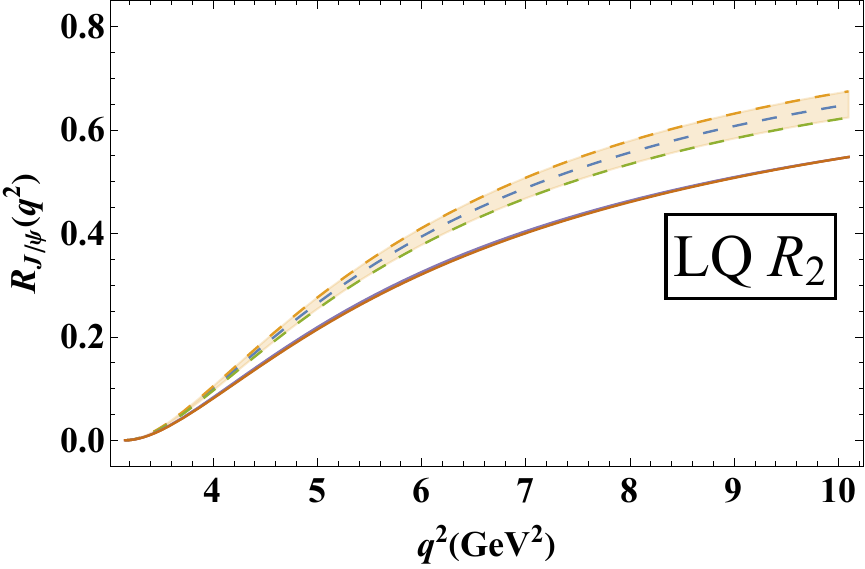}
		\includegraphics[scale=0.39]{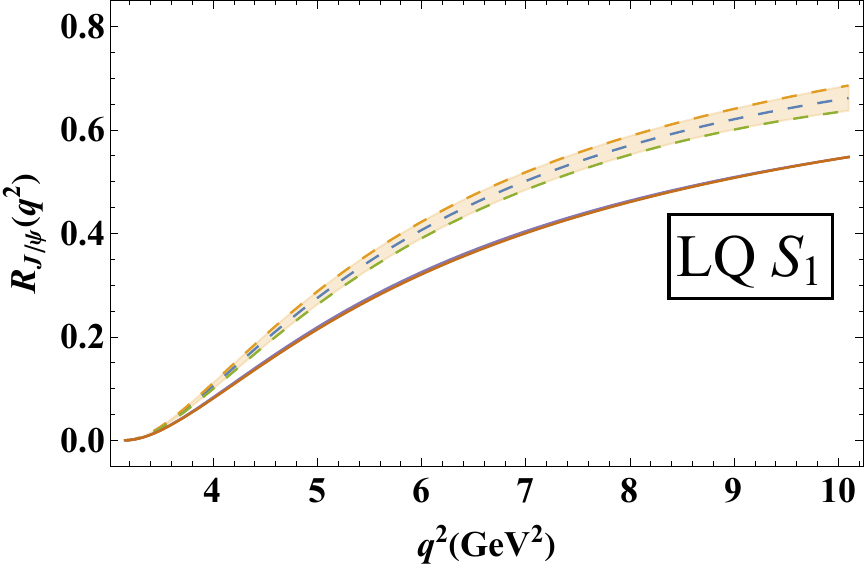}
		\includegraphics[scale=0.39]{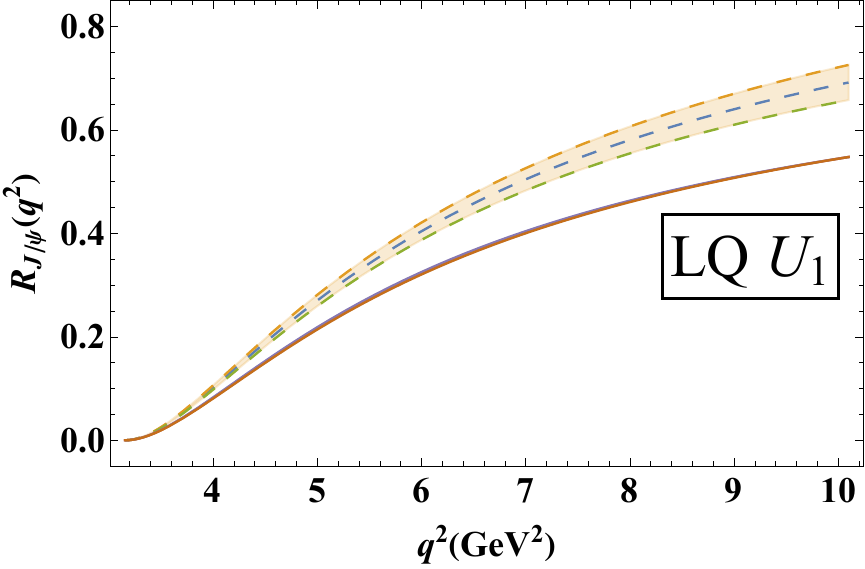}
		\caption{
			Predictions for the differential ratio $R_{J/\psi}(q^2)$ in the SM (red solid lines) and in the leptoquark models (blue dashed lines), corresponding to the best-fit Wilson coefficients. The bands include errors from form factors (for SM and leptoquark models) and Wilson coefficients (leptoquark models).}
		\label{fig:lpRq2}
	\end{center}
\end{figure}

\begin{figure}
	\begin{center}
        \includegraphics[scale=0.39]{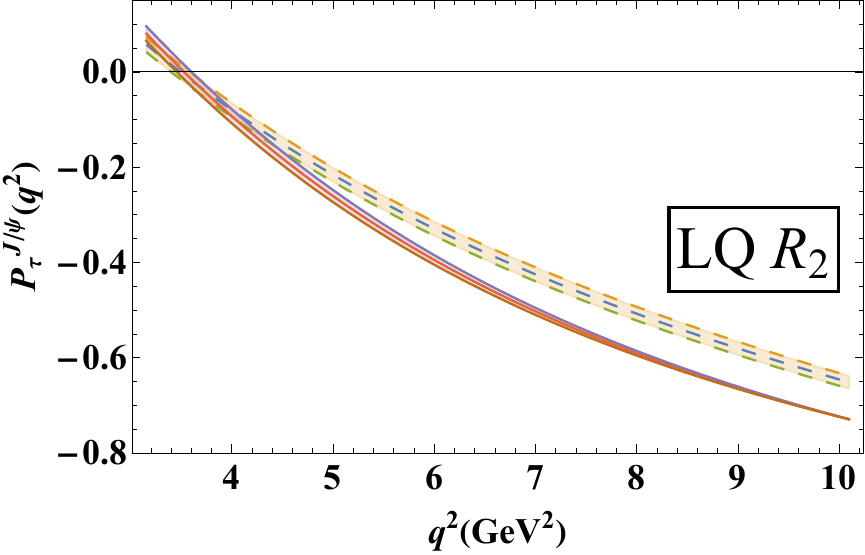}
		\includegraphics[scale=0.39]{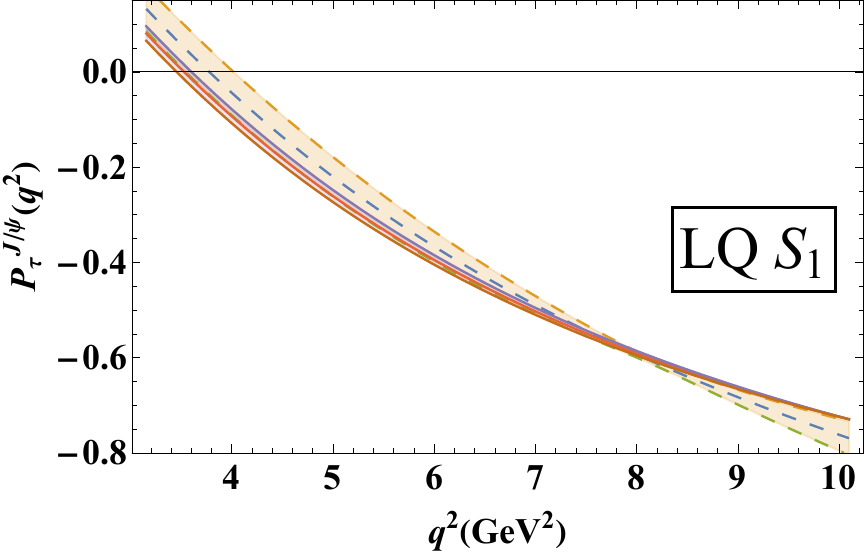}
		\includegraphics[scale=0.39]{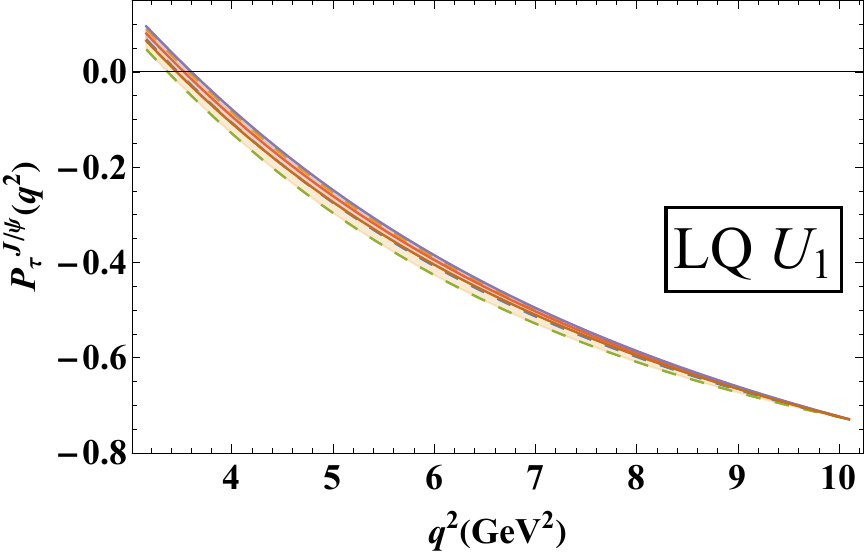}
		\caption{
			Predictions for the differential polarization fraction $P_\tau^{J/\psi}(q^2)$ in the SM (red solid lines) and in the leptoquark models (blue dashed lines), corresponding to the best-fit Wilson coefficients. The bands include errors from form factors (for SM and leptoquark models) and Wilson coefficients (leptoquark models).}
		\label{fig:lpPtauq2}
	\end{center}
\end{figure}

\begin{figure}
	\begin{center}
        \includegraphics[scale=0.39]{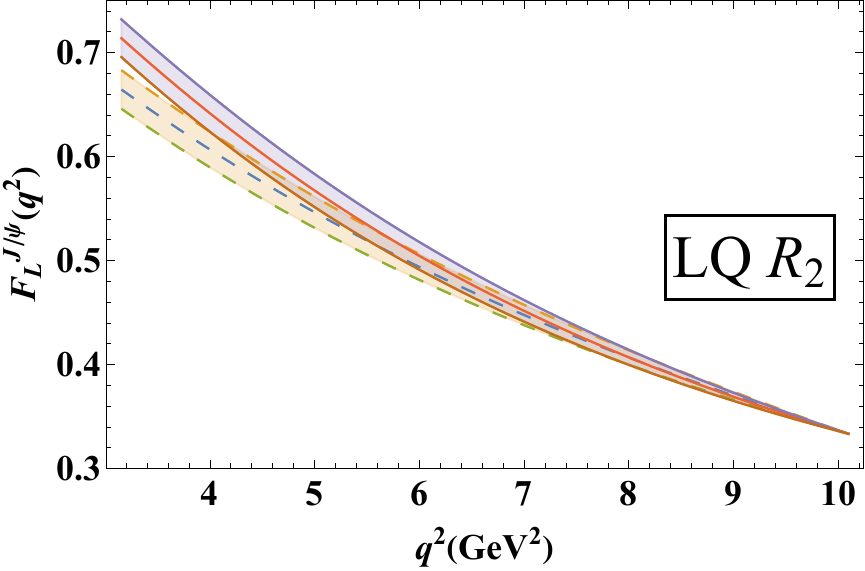}
		\includegraphics[scale=0.39]{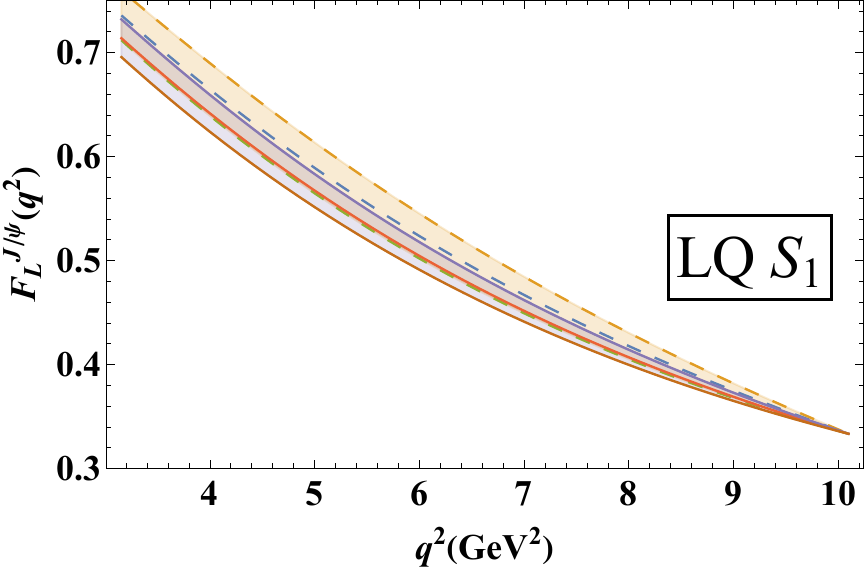}
		\includegraphics[scale=0.39]{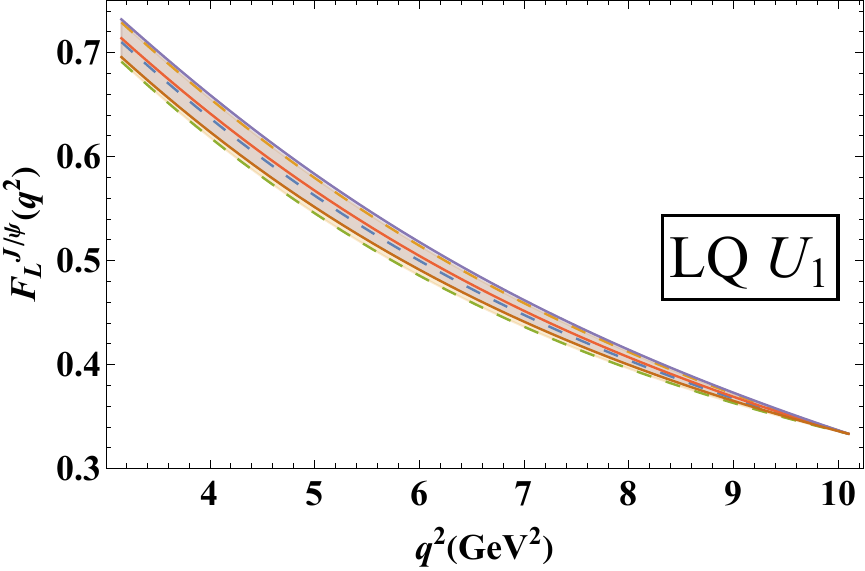}
		\caption{
			Predictions for the differential polarization fraction $F_L^{J/\psi}(q^2)$ in the SM (red solid lines) and in the leptoquark models (blue dashed lines), corresponding to the best-fit Wilson coefficients. The bands include errors from form factors (for SM and leptoquark models) and Wilson coefficients (leptoquark models).}
		\label{fig:lpPq2}
	\end{center}
\end{figure}

\begin{figure}
	\begin{center}
        \includegraphics[scale=0.39]{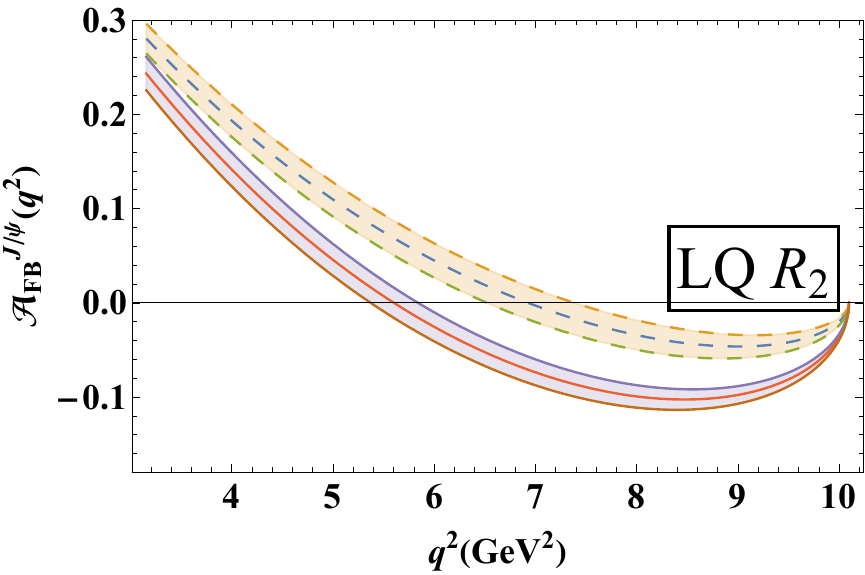}
		\includegraphics[scale=0.39]{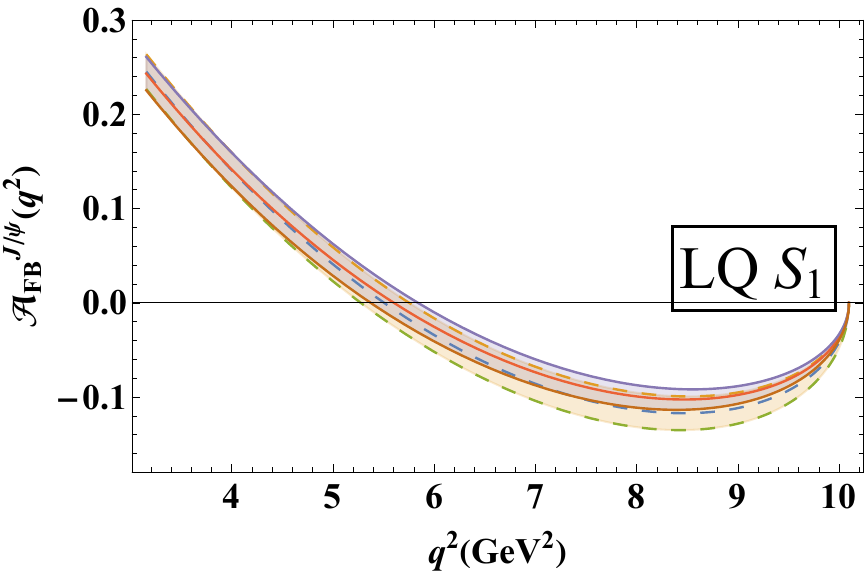}
		\includegraphics[scale=0.39]{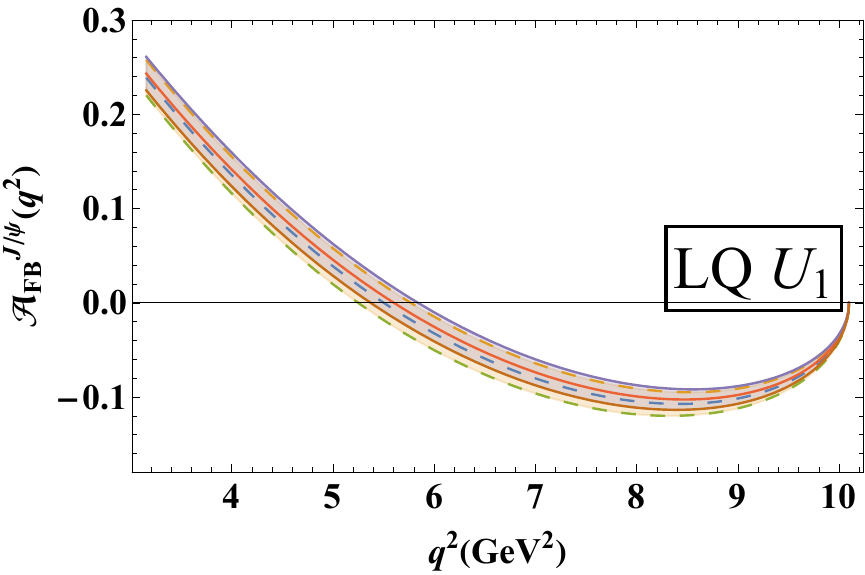}
		\caption{Predictions for the differential forward-backward asymmetry $\mathcal A_{FB}^{J/\psi}(q^2)$ in the SM (red solid lines) and in the leptoquark models (blue dashed lines), corresponding to the best-fit Wilson coefficients. The bands include errors from form factors (for SM and leptoquark models) and Wilson coefficients (leptoquark models).}
		\label{fig:lpAFB}
	\end{center}
\end{figure}
\section{Summary and conclusions}\label{sec:SUM}
The discrepancy between experimental measurements and theoretical predictions for the $b\to c\tau\nu$ transition continues to draw attention in the community. Future measurements on channels complementary to $B\to D^{(*)}\tau\nu$ will be important for the investigation of new physics/LFUV. Such channels certainly include $B_c\to J/\psi\tau\nu$ for which the measurement of the LFU ratio $R(J/\psi)$ has been done by LHCb. In addition, for the $\Lambda_b\to\Lambda_c\tau\nu$ transition, more recently $R(\Lambda_c)$ has also been measured by LHCb, which is consistent with the SM prediction at $1\sigma$, although the error is still sizable.

In this work, we have conducted an analysis for the $B_c\to J/\psi\tau\nu$ channel by employing the lattice results on the (axial-)vector $B_c\to J/\psi$ form factors and taking into account the recently measured $R(\Lambda_c)$ in the global fit. We have studied the new physics effects in   a model independent manner and in the leptoquark models including $R_2$, $S_1$ and $U_1$, which can address the $b\to c\tau\nu$ anomalies. In order to study the tensor new physics operator as well as the scalar leptoquark models $R_2$ and $S_1$, we have determined the $B_c\to J/\psi$ tensor form factors using the lattice data on the (axial-)vector form factors and the NRQCD relations including the relativistic corrections from the valence quarks between (axial-)vector and tensor form factors.

Based on the lattice QCD+NRQCD $B_c\to J/\psi$ form factors, we have performed two sets of fits of the Wilson coefficients in the weak effective theory and new physics couplings in the leptoquark models to the $b\to c\tau\nu$ data, with and without inclusion of $R(\Lambda_c)$. By imposing the relaxed constraint $\mathcal B(B_c\to\tau\nu)<30\%$ following the recent study of LEP1 data and $B_c$ lifetime, we have found that $V_1$, $V_2$ and $T$ scenarios are the favoured single operator scenarios, while $S_1$ and $S_2$ scenarios are excluded by the $b\to c\tau\nu$ measurements and the limit on $\mathcal B(B_c\to\tau\nu)$, respectively. By adding the data on $R(\Lambda_c)$ in the fit, which shows a different tendency in the deviation from the SM, $\chi^2/d.o.f.$ for all scenarios increase, and the $R_2$ leptoquark model turns out to have the best fit results with the smallest $\chi^2/d.o.f.$ among all new physics scenarios/models which we have considered.

Furthermore, we have also used the form factors and the Wilson coefficients/leptoquark couplings to study the new physics effects. The effects of the $B_c \to J/\psi$ form factors have been found tiny on the global fit to the $b\to c\tau\nu$ data but visible on the predictions for other observables we have considered, including the integrated observables $R(J/\psi)$, $P_\tau(J/\psi)$, $F_L(J/\psi)$ and $\A_{FB}(J/\psi)$, and the corresponding $q^2$ dependent observables. Among the observables, we have found that $P_\tau(J/\psi)$ in $R_2$ leptoquark model, and $\A_{FB}(J/\psi)$ in $V_2$ scenario and $R_2$ leptoquark model are distinct from the SM predictions. Besides, all observables in $T$ scenario also deviate from the SM values, although they are less distinguishable due to large errors. Future measurements of these observables will shed light on the possible new physics effects on the $B_c\to J/\psi\tau\nu$ decay.
\ack
The authors would like to thank Ji-Bo He and Wen-Qian Huang for useful discussions. This work is partly supported by the National Natural Science Foundation of China with Grant No.12070131001, No.12075124 and the National Key Research and Development Program of China under Contract No.2020YFA0406400. The research is partly supported by an appointment to the YST Program at the APCTP through the Science and Technology Promotion Fund and Lottery Fund of the Korean Government, the Korean Local Governments - Gyeongsangbuk-do Province and Pohang City.
\section*{References}
\bibliography{myreference}
\end{document}